\documentclass{aa}  
\usepackage{graphicx}
\usepackage{txfonts}
\usepackage{lipsum}
\usepackage[colorlinks=true,
            linkcolor=black,
            citecolor=blue,
            urlcolor=black]{hyperref}
\usepackage{subcaption}
\usepackage{lscape}
\usepackage{placeins}

\newcommand{\BB}{{\cal B}}

\begin{document}

\title{Polarized emission of orbiting hot-spots near Sagittarius A*:\\  effects of electromagnetic interaction}
\titlerunning{Polarized emission of orbiting hot-spots}
\authorrunning{Tlemissov et al.}

\author{
  Abylaikhan Tlemissov\inst{1},
  Arman Tursunov\inst{1} 
  \and
  Maciek Wielgus\inst{2}
}

\institute{
  Research Centre for Theoretical Physics and Astrophysics, Institute of Physics, 
  Silesian University in Opava, \\
  Bezru\v{c}ovo n\'{a}m.~13, CZ-74601 Opava, Czech Republic\\
  \email{tle0002@slu.cz; arman.tursunov@physics.slu.cz}
  \and
  Instituto de Astrofísica de Andalucía-CSIC, 
  Glorieta de la Astronomía s/n, E-18008 Granada, Spain\\
  \email{maciek.wielgus@gmail.com}
}

\abstract
 {We investigate the polarimetric signatures of orbiting hot-spots around a Schwarzschild black hole in the presence of an external magnetic field, accounting for the electromagnetic interaction between the charged emitter and the field. Using a general-relativistic model that incorporates synchrotron emission and ray-tracing of light propagation, we analyze how the electromagnetic interaction parameter modifies the observed polarization patterns, with particular emphasis on the behavior of the electric vector position angle (EVPA) and the time-evolving polarization loops in the $Q$-$U$ plane. Applying the model to millimeter wavelength ALMA observations of Sagittarius~A*, we explore the parameter space that best reproduces the asymmetry, time ratio, and area ratio of the observed polarization loops. We find that the inclusion of a small positive interaction parameter increases the symmetry of the loops and 
the time ratio, while a negative magnetic parameter introduces strong asymmetry and fails to reproduce the data. Our results indicate that electromagnetic interaction can lead to ambiguity in the estimation of the system parameters such as orbital inclination or hot-spot velocity.}


\maketitle
\nolinenumbers
   
\section{Introduction} \label{sec:intro} 

Polarimetric observations of black holes provide crucial information on the extreme gravitational and electromagnetic environments in their vicinity. Recent groundbreaking observations by the Event Horizon Telescope (EHT) Collaboration at 230 GHz have opened new avenues for probing the complex interplay between gravity and magnetic fields around supermassive black holes (SMBHs). 
Specifically, polarization images of the SMBH M87* \citep{2021ApJ...910L..12E} and Sagittarius A* \citep[Sgr~A*;][]{2024ApJ...964L..25E} have unveiled structures suggesting presence of predominantly poloidal magnetic field components, especially in the latter case \citep{EHT_SgrA_P8}. Furthermore, observations by the 
\cite{2018A&A...618L..10G,2023A&A...677L..10G} in near infrared have detected distinct polarization loops associated with orbiting flaring matter, offering additional constraints on the dynamics and magnetic field configurations close to an event horizon \citep{Michi2020,Alejandra2020,2024A&A...691A.327Y}. 

Complementing these observations, \cite{2022A&A...665L...6W} reported detection of polarization loops by the Atacama Large Millimeter/submillimeter Array (ALMA), revealing time-dependent polarization signatures associated with orbital motion close to Sgr~A*, following a high energy flare observed by Chandra \citep{2022ApJ...930L..13E,Wielgus2022_lightcurves}.
These polarization loops seen as cyclic trajectories in the Stokes $Q$–$U$ plane of linear polarization, provide an evidence of transient, localized emission components orbiting in the innermost region near the event horizon \citep[see also][]{Vincent2024}. By analyzing these loops, the authors constrained the orbital properties of the emitting regions, such as their orbital radii and periods, revealing details about plasma dynamics and magnetic field geometry close to Sgr~A*. This observational finding significantly enhances our understanding of Sgr~A* black hole and its astrophysical environment, offering novel observational constraints on theoretical models \citep{2024A&A...685A.142Y,Levis2024}.

Despite these advances, interpreting polarization signals from near-horizon regions remains challenging due to uncertainties in the plasma motion and magnetic field structure. One key source of uncertainty arises from the assumption of Keplerian motion for emitting material around black holes, which might not hold true in strong magnetic field environments. In reality, the emitting plasma consists of charged particles whose motion can deviate significantly from purely gravitational trajectories due to electromagnetic interactions with vertical magnetic fields \citep{2020ApJ...897...99T}. Such interactions can introduce substantial non-Keplerian effects, potentially mimicking observational signatures attributed to black hole spin or other gravitational phenomena \citep[see, e.g., appendix C from][]{2022A&A...665L...6W}.

EHT observations consistently support the presence of dominant vertical magnetic fields, with significantly weaker toroidal or radial components, reinforcing the importance of vertical magnetosphere models in describing the Galactic Center environment \citep{2015Sci...350.1242J,2024MNRAS.532.1522J}. These vertical and large-scale magnetic fields play a crucial role in particle acceleration, jet formation, and overall energy extraction processes in black hole systems \citep{1977MNRAS.179..433B,2019Univ....5..125T}.

Motivated by these observations, we propose an extension of the polarized synchrotron emission model by \cite{2021ApJ...912...35N} and \cite{2021PhRvD.104d4060G} from weakly charged hot-spots orbiting a Schwarzschild black hole embedded in an vertical magnetic field. In this model, we explicitly account for the electromagnetic interaction between the potentially charged emitting hot-spot and the magnetic field, introducing a dimensionless magnetic interaction parameter.

Standard GRMHD simulations typically assume exact charge neutrality at fluid scales. In reality, a magnetized plasma is only \emph{quasi}-neutral: a tiny excess of charge carriers can exist without violating global neutrality. The size of this excess is set by electrostatic screening. 
In a quasi-neutral plasma, the fractional charge imbalance required to sustain a smooth electrostatic potential variation $\Delta \Phi$ across a region of size $L$ can be estimated as \citep[see, e.g.,][]{HazeltineWaelbroeck2004,2016ippc.book.....C} 
\begin{equation}
\frac{n_q}{n} 
\approx 
\lambda_D^2 \nabla^2 \frac{e \Phi}{k_B T} 
\approx \left(\frac{\lambda_D}{L}\right)^2\,\frac{e\,\Delta\Phi}{k_B T},
\label{eq:q-neutrality}
\end{equation}
where $n_q\equiv n_i-n_e$ is the net charge number density, $n$ is a  total  number density ($n\simeq n_e\simeq n_i$), $T$ is a plasma temperature, and $\lambda_D^2\equiv k_B T/(4 \pi  n e^2)$ is the Debye length. 
Eq.~\eqref{eq:q-neutrality} is derived by using Poisson’s equation $\nabla^2\Phi=- 4 \pi e n_q $ with the scale estimate $\nabla^2\Phi\sim\Delta\Phi/L^2$.

Such charge imbalances can naturally be generated by the relativistic motion of plasma around a black hole in the presence of an ordered magnetic field approximately orthogonal to the orbital plane, which induces charge separation and a nonzero net charge density. In pulsar magnetospheres the same special–relativistic effect produces the well-known Goldreich--Julian charge density \citep{1969ApJ...157..869G,1992MNRAS.255...61M}. 
Neglecting general relativistic corrections for the moment, Maxwell’s equations in the lab frame together with Ohm’s law imply that the charge density required to reduce the co-moving electric field is 
\begin{equation}
\rho_q = \frac{1}{4\pi}\nabla \cdot \mathbf{E}
= -\frac{1}{4\pi c} \, \nabla \cdot(\mathbf{v}\times\mathbf{B})
\approx -\frac{\boldsymbol{\Omega}\!\cdot\!\mathbf{B}}{2\pi c},
\end{equation}
for rigid rotation with the speed   $\mathbf{v}=\boldsymbol{\Omega}\times\mathbf{r}$ and a vertical field $\mathbf{B}$. For a hot-spot orbiting Sgr~A* near the innermost stable circular orbit (ISCO) of non-rotating black hole and a vertical field of strength $B\approx 10$ G \citep{EHTC_SgrA_p5,EHT_SgrA_P8}, the corresponding excess charge number density is of the order of 
\begin{equation}
n_{q}\equiv\frac{|\rho_q|}{e}\;\approx 
10^{-4}
\left(\frac{B}{10\,{\rm G}}\right) 
\left(\frac{\Omega}{\Omega_{\rm ISCO}}\right)\ {\rm cm^{-3}}. 
\label{eq:nq}
\end{equation}
Given the total plasma number density associated with the flare components, $n \sim 10^{7\pm 1}\ \mathrm{cm^{-3}}$ \citep{2003ApJ...598..301Y,2012A&A...537A..52E,2024A&A...685A.142Y}, the ratio of excess to total carriers is $n_q/n \sim 10^{-11\pm 1}$. This means that only about one particle in every $10^{10}-10^{12}$ contributes to the net charge, underscoring that the plasma remains overwhelmingly neutral even when such charges are dynamically relevant.

One can show that the right-hand side of \eqref{eq:q-neutrality} yields a similar order-of-magnitude estimate. For a hot-spot with characteristic size $L \sim r_g = GM/c^2 \simeq 6\times10^{11}$ cm and plasma temperature $T\simeq 10^{11}$ K, moving near the ISCO, the inductive electric field has a magnitude $E\simeq (v/c) B$. This corresponds to a potential difference across the hot-spot of order  $\Delta \Phi \sim E L = (v/c) B L$. 
Substituting this into \eqref{eq:q-neutrality}, we obtain a charge imbalance of order $|n_q|/n \sim 10^{-10\pm1}$. 

Although this fraction is vanishingly small, the integrated excess charge over a hot-spot of size $L\sim r_g$ corresponds to a macroscopic total charge, $Q\simeq e\,n_{q}\, L^3 \sim 10^{12}$C. Such a charge is large enough for its electrostatic and magnetic coupling to the ambient field, potentially modifying its orbital motion and the polarization signature of its emission. General–relativistic extensions of the above estimates \citep[see, e.g.,][]{1992MNRAS.255...61M,1982MNRAS.198..339T,2020ApJ...897...99T} yield comparable magnitudes. 
To estimate the dynamical effect of the excess charge, we define a dimensionless parameter characterizing the strength of the electromagnetic interaction between the charged hot-spot and the ambient magnetic field

\begin{equation}
\mathcal{B} = \frac{Q B}{m_{hs}} \left(\frac{G M}{c^4}\right)\equiv\frac{n_{q}}{n}\times\frac{eB}{m_{i}}\left(\frac{G M}{c^4}\right),
\label{eq:BBparam}
\end{equation}

where $Q$ and $m_{hs}$ are the total charge and mass of the hot-spot. This parameter represents the ratio of the characteristic Lorentz force to the gravitational force, and governs deviations from purely geodesic motion.
For a characteristic ratio $n_{q}/n \sim 10^{-11}-10^{-9}$ we obtain $\mathcal{B} \sim 10^{-5}-10^{-3}$ for an $e^{-}$--p plasma and $\mathcal{B} \sim 10^{-2}-1$ for an $e^{-}$--$e^{+}$ plasma. The upper end of the range  
corresponds to the pair-dominated plasma inertia together with the largest assumed charge imbalance ($n_q/n$). For smaller charge imbalance or in an electron--proton plasma, substantially lower values of $\mathcal{B}$ are expected. 
The allowed range of this parameter for Sgr~A* has been constrained by \citet{2020ApJ...897...99T}; see their Eq.~(33), which yields $10^{-5} < |\mathcal{B}| < 10$, for 10 G magnetic field near Sgr A*. The sign of the parameter $\mathcal{B}$ depends on the sign of the hot-spot charge and the relative orientation between the magnetic field and the orbital motion, and can be either positive or negative \citep{2010PhRvD..82h4034F,2015CQGra..32p5009K}.

Charge separation is also driven by general relativity itself: frame dragging twists magnetic field lines around a spinning black hole and induces an electric field.  \cite{1974PhRvD..10.1680W} showed that a Kerr black hole in a uniform magnetic field acquires an induced charge $Q_{\rm W}$ set by the spin and the vertical magnetic field. For Sgr~A* this constrains the black hole charge to $|Q_{\rm BH}|\!\lesssim\!10^{15}$\,C \citep{2018MNRAS.480.4408Z}. Classic analyses of the coupled black hole–magnetosphere system by \cite{1975PhRvD..12.2959R} show that, for a wide class of stationary accretion configurations, the global charge of the magnetosphere balances that of a black hole, $Q_{\rm mag} \approx - Q_{\rm BH}$, ensuring overall neutrality of the system. These estimates show the significance of electromagnetic interactions: even a tiny deviation from charge neutrality, while still maintaining quasi-neutrality, can sustain dynamically relevant electric fields in the magnetized vicinity of Sgr~A*.

In this paper we systematically explore how variations in the electromagnetic interaction parameter can influence polarization signatures discussed by \cite{2022A&A...665L...6W}.  Further, we compare our theoretical predictions with ALMA observations of polarization loops from Sgr A*, focusing on the effect of the magnetic interaction on the $QU$-pattern parameters considered by \cite{2022A&A...665L...6W} and \cite{2022A&A...668A.185V}, namely the degree of asymmetry of the loop, ratio of the full loop period to inner loop duration and ratio of the areas of large and small loops on the $Q$–$U$ plane. The introduced magnetic field interaction parameter $\mathcal{B}$ can also be regarded as an effective parameter controlling deviations from the geodesic motion.

The paper is organized as follows. We start from the theoretical description of the dynamics of a charged emitter in a Schwarzschild spacetime in the presence of vertical magnetic field given by the exact solution of \cite{1974PhRvD..10.1680W}. We then calculate the polarization vectors, radiative transfer equations for propagation of the polarized light, and derive the observable parameters in the case of a charged emitter. 
We then apply our model to the ALMA polarization loops of Sgr A* \citep{2022A&A...665L...6W} to constrain the magnetic interaction parameter and assess its impact on orbital dynamics and the interpretation of observational data. 

\vspace{0.5cm}

\section{Orbiting hot-spot with electromagnetic interaction}

We start from introducing line element in a spherically symmetric Schwarzschild spacetime
\begin{equation}
    ds^2=-f(r)dt^2+\frac{dr^2}{f(r)}+r^2(d\theta^2+\sin^2{\theta}d\phi^2),
\end{equation}
where $f(r)=1-2M/r$ is a lapse function, $M$ is the mass of the central object in geometric units. Without loss of generality, we set the black hole mass to unity, $M=1$, in the following. The mass will be reinstated later when expressing equations in astrophysically relevant units. 

\cite{1974PhRvD..10.1680W} introduced a method to generate test electromagnetic fields in stationary, axisymmetric spacetimes with Killing vector symmetries, obtaining a solution
\begin{equation}    A^{\mu}=C_{t}\xi_{t}^{\mu}+C_{\phi}\xi_{\phi}^{\mu},
\end{equation}
where the constants $C_{t}$	and $C_{\phi}$ are determined by the asymptotic behavior of the field at infinity. In the case of non rotating and uncharged black hole, the only relevant Killing vector is the rotational one $\xi_{\phi}=\partial/\partial \phi$. Therefore, assuming asymptotic magnetic field $B$ strength, resulting vector potential reads
\begin{equation}
 A_{\phi}=\frac{1}{2}Br^2\sin^2{\theta} \, ,
\end{equation}
with the antisymmetric electromagnetic field tensor components given by 
\begin{equation}
\label{fied_tensors}
    F_{r\phi}=B r \sin^2{\theta}, \quad F_{\theta \phi}=B r^2 \cos{\theta}\sin{\theta}
\end{equation}
The Hamiltonian governing the motion of a charged hot-spot can be expressed as  \citep{1984ucp..book.....W,2015CQGra..32p5009K} 
\begin{equation}
H_{p} = \frac{1}{2} g^{\alpha\beta} \left(\pi_{\alpha} - q A_{\alpha}\right) \left(\pi_{\beta} - q A_{\beta}\right) + \frac{1}{2} m^2,
\end{equation} 
where
$m$ is the mass of the charged particle and 
$q$ is its electric charge. The generalized canonical and kinematical four-momenta are defined, respectively 
\begin{equation}
\pi^{\mu} = p^{\mu} + q A^{\mu} \,, \quad
p^{\mu} = m u^{\mu} \, .
\end{equation}
Due to symmetries of the background spacetime and the external magnetic field, one can introduce conserved quantities: the energy and the axial component of angular momentum 
\begin{eqnarray}
    E&=&-\pi_{t}=m f(r)\frac{dt}{d\tau},\\
    L&=&\pi_{\phi}=mr^2\sin^2{\theta}\left(\frac{d\phi}{d\tau}+\frac{qB}{2m}\right)
    \label{conserved_quantity}
\end{eqnarray}
For the convenience we use dimensionless magnetic parameter $\mathcal{B}$, introduced in (\ref{eq:BBparam}), and introduce specific energy $\mathcal{E}$ and the specific angular momentum $\mathcal{L}$ as follows 
\begin{equation} \label{eq:param}
    \mathcal{B}=\frac{qB}{2m},\quad \mathcal{E}=\frac{E}{m},\quad \mathcal{L}=\frac{L}{m} \,. 
\end{equation} 
Using the above quantities, one can rewrite the Hamiltonian as
\begin{equation}
    H=\frac{1}{2}f(r)p_{r}^2+\frac{1}{2r^2}p_{\theta}^2+\frac{m^2}{2f(r)^2}\left(V_{\rm{eff}}-\mathcal{E}^2\right) \, ,
\end{equation}
where the effective potential is taken in the form 
\begin{equation}    V_{\rm{eff}}=f(r)\left[1+\left(\frac{\mathcal{L}}{r\sin{\theta}}-\mathcal{B}r\sin{\theta}\right)^2\right] \, .
\end{equation} 
The effective potential, representing energetic boundary of the hot-spot, is symmetric under the transformation $\left(\mathcal{L},\mathcal{B}\right)\rightarrow \left(-\mathcal{L},-\mathcal{B}\right)$ which implies two equivalent scenarios: either $\mathcal{L}>0,$ $\mathcal{B}<0$ is equivalent to $\mathcal{L}<0,$ $\mathcal{B}>0$ or $\mathcal{L}>0,$ $\mathcal{B}>0$ is equivalent to $\mathcal{L}<0,$ $\mathcal{B}<0$. We choose the positive angular momentum of the charged hot-spot to represent the clockwise motion, which is supported by observations \cite{2018A&A...618L..10G, Wielgus2022_lightcurves}.

Depending on the sign of $\mathcal{B}$, one can distinguish two qualitatively distinct cases, differing in the orientation of the effective Lorentz force: for $\mathcal{B}<0$ the Lorentz force is attractive, acting towards the black hole, while for $\mathcal{B}>0$ it is repulsive, acting outwards from the black hole. In analogy with the prograde and retrograde motion in Kerr spacetime, the two above cases in Schwarzschild spacetime can be referred to as Larmor ($\mathcal{B}<0$) and anti-Larmor ($\mathcal{B}>0$) motions, defined with respect to the magnetic field orientation. 

The extrema points of the effective potential give locations of circular orbits. For a given radius $r$ the corresponding specific angular momentum and energy of a hot-spot at circular orbit is given by 

\begin{eqnarray}   
(\mathcal{L}_{\rm c.o.})_{\pm}&=&\frac{-\mathcal{B}r^2\pm rF}{r-3} \, , \\
(\mathcal{E}_{\rm c.o.})_{\pm}&=&\left(f(r)\left[1+\left(\frac{(\mathcal{L}_{\rm{c.o}})_{\pm}}{r}-\mathcal{B}r\right)^{2}\right]\right)^{1/2} \, ,
\end{eqnarray} 

where
\begin{equation}
    F(r;\mathcal{B})=\sqrt{\mathcal{B}^2r^2(r-2)^2+r-3}\,.
\end{equation}
Thus, using \eqref{conserved_quantity} we find the angular velocity of at the circular orbit  

\begin{equation}
    \Omega_{\pm}=\frac{f(r)}{\mathcal{E}_{\rm c.o.}}\left(\frac{\mathcal{L}_{\rm c.o.}}{r^2}-\mathcal{B}\right)
    \label{ang_vel}
\end{equation} 

where the $\pm$ sign corresponds to the two branches of motion, in analogy with prograde and retrograde orbits in Kerr spacetime, here referred to as Larmor and anti-Larmor cases.  
However, within a narrow radial range $2<r<3$, both branches of the solution reduce to the same orbital orientation, which can be either Larmor-like or anti-Larmor-like, depending on the sign of the product ${\cal L} \,\BB$,  see details in \cite{2010PhRvD..82h4034F,2015CQGra..32p5009K,2016PhRvD..93h4012T}. With that, the circular motion under $r<3$  corresponds to the case with the repulsive Lorentz force only. 
\begin{figure}[htbp]
    \centering
    \includegraphics[width=0.8\linewidth]{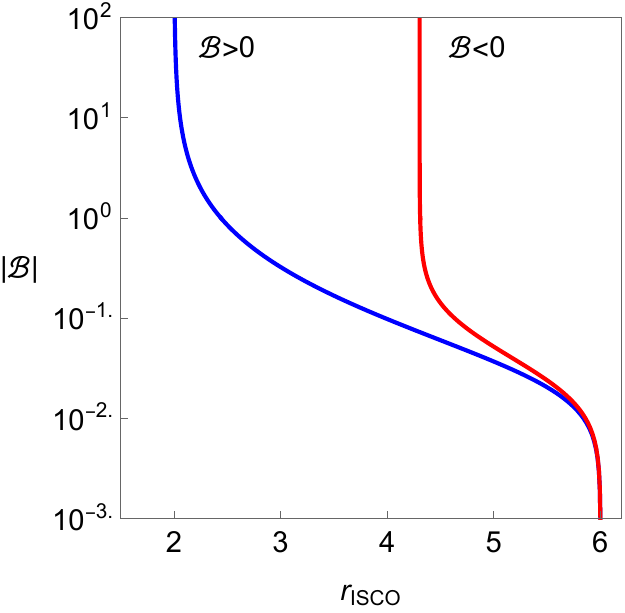}
    \caption{Dependence of the ISCO radius $r_{\rm{ISCO}}$ on the magnetic field interaction parameter $\mathcal{B}$ in the presence of a uniform magnetic field. In both cases of positive and negative $\mathcal{B}$, the ISCO radius decreases with increasing $\BB$.}
    \label{fig:ISCO}
\end{figure}
The local radial extrema of $\mathcal{L}_{\rm c.o.}$ correspond to the ISCO, defining their radius, angular momentum, and energy.
The ISCO condition reads 
\begin{equation}
\mathcal{B}^2 r \left(2r^3 - 9r^2 + 8r - 12\right) - \mathcal{B} (r - 6) Y + r/2 - 3 = 0,  
\end{equation}%
where 
\begin{equation}
Y(r; \mathcal{B}) = \sqrt{\mathcal{B}^2 r^2 \left(5r^2 - 4r + 4\right) + 2r}.
\end{equation}
The behavior of ISCO with respect to magnetic field interaction parameter is shown in Fig. \ref{fig:ISCO}. In the absence of electromagnetic interaction ($\BB = 0$), the ISCO is located at the standard Schwarzschild spacetime value of $r = 6\,r_{g}$.

\begin{figure*}
\centering
\includegraphics[width=0.9\textwidth]{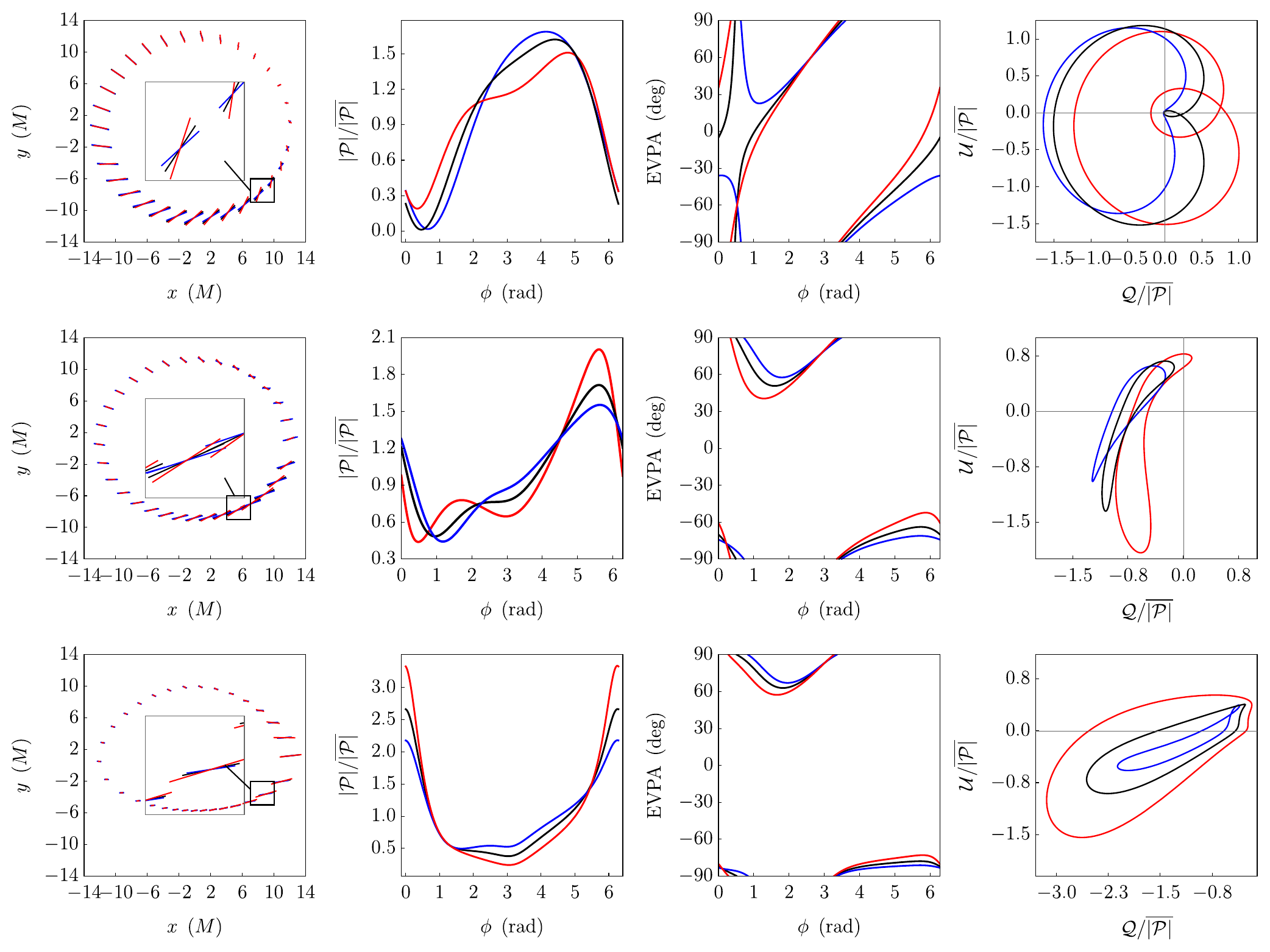}
\caption{From left to right: polarization patterns, polarization intensity values, EVPA, and $Q$-$U$ fluxes shown in the case of $r_{s}=11$, $\chi=-\pi/2$, $\beta=\beta_{R}$ and for different inclination angles: $20^{\circ},\ 40^{\circ},\ 60^{\circ}$ (from top to bottom), with curves corresponding to different magnetic field parameter: $\mathcal{B} = 0.01$ (blue), $\mathcal{B} = 0$ (black), and $\mathcal{B} = -0.01$ (red). In each polarization panel, the polarization segment lengths are normalized relative to the maximum intensity. The intensity and Q--U fluxes are normalized by the average polarization magnitude. This average can be defined as an angular average $ \overline{|\mathcal{P}|} = \tfrac{1}{2\pi}\int_{0}^{2\pi} |\mathcal{P}(\phi)|\, d\phi $. Since the angular frequency $\omega$ is constant in time, the same quantity can equivalently be expressed as a time average over one period $ T=2\pi/\omega $, namely $ \overline{|\mathcal{P}|} = \tfrac{1}{T}\int_{0}^{T} |\mathcal{P}(\omega t)|\, dt $.
}
\label{polarimatic_image_pic}
\end{figure*}

\begin{figure*}
\centering
\includegraphics[width=0.9\textwidth]{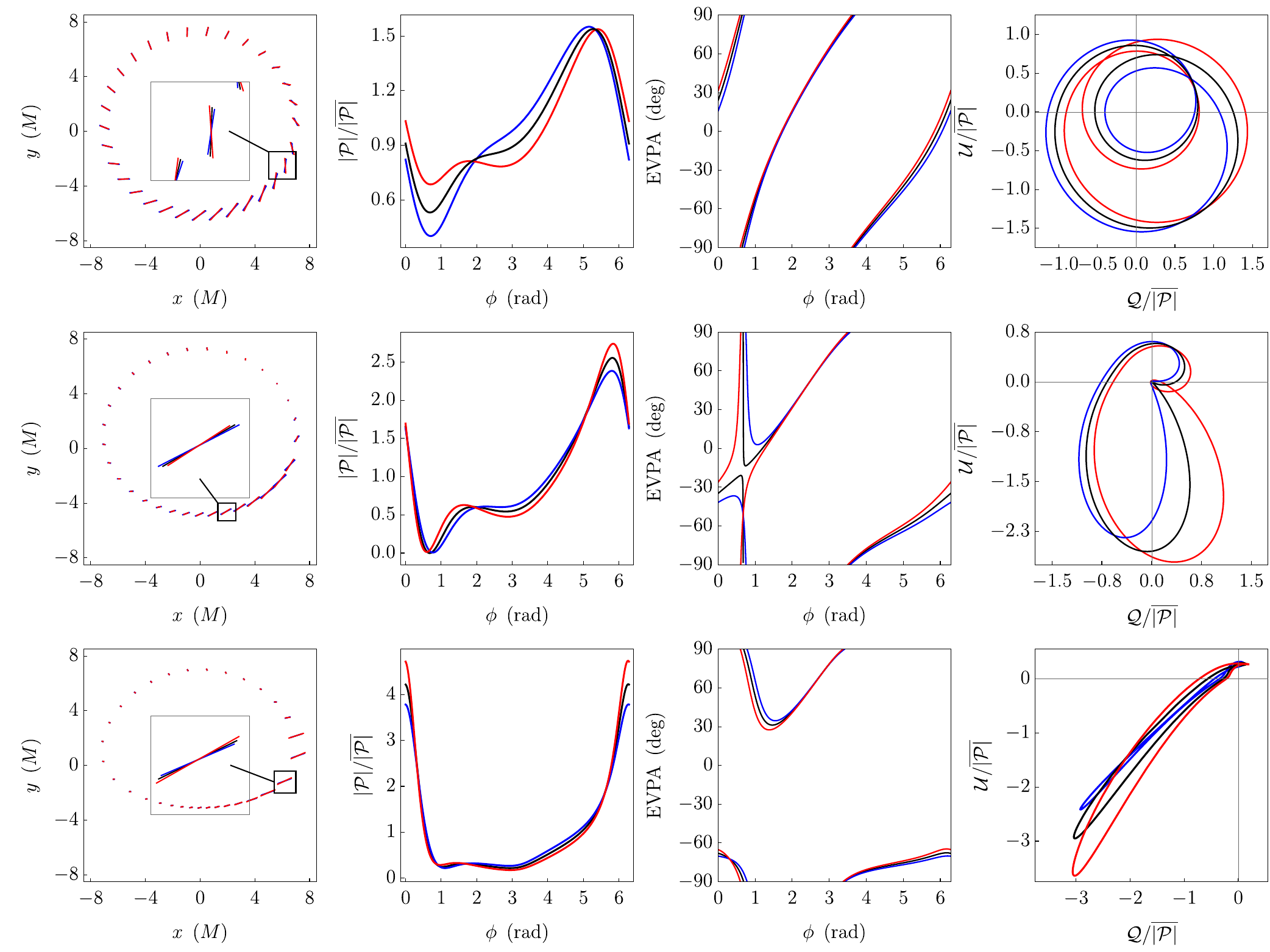}
\caption{Same as Figure \ref{polarimatic_image_pic}, but for $r_s=6$. Strong gravity effects are more prominent. 
}
\label{polarimatic_image_pic2}
\end{figure*}

\section{Polarized signatures of hot-spots} 

\subsection{Polarization formalism 
} \label{sec:polarization}

In this section we discuss the key aspects of the polarization of light around black hole within a semi-analytic 
model introduced by \citet{2021ApJ...912...35N} and \citet{2021PhRvD.104d4060G}, and modified here by the inclusion of electromagnetic interaction.  
In this model, radiation is emitted by the fluid from the equatorial plane $\theta_{s}=\pi/2$, confined to a point-like emitter located at the dimensionless radius $r_{s} \equiv r c^2/GM$ in Schwarzschild coordinates.  The orbital plane of the hot-spot is inclined at an angle $\theta_{o}$ relative to a face-on orientation, as seen by a distant observer.

Below, we transform the components of vectors and tensors into the local orthonormal frame at the arbitrary point $P$ (hereafter, $P$-frame; the geometry of the different reference frames is described in more detail in \citet{2021ApJ...912...35N}) in the equatorial plane via 

\begin{equation}
    V_{(a)}=e^{\mu}_{(a)}V_{\mu} \, , \quad T_{(a)(b)}=e^{\mu}_{(a)}e^{\nu}_{(b)}T_{\mu\nu} \, ,
\end{equation}

where in Schwarzschild spacetime 

\begin{equation}
 e^{\mu}_{(a)} =
\begin{pmatrix}
f(r)^{-\frac{1}{2}} & 0 & 0 & 0 \\
0 & f(r)^{\frac{1}{2}} & 0 & 0 \\
0 & 0 & r^{-1} & 0 \\
0 & 0 & 0 & ({r\sin{\theta})^{-1}}
\end{pmatrix} \,\,.
\end{equation}

For the chosen magnetic field configuration given by \eqref{fied_tensors}, the components of the magnetic field in the $P$-frame are:

\begin{eqnarray}
    F_{\hat{\theta}\hat{\phi}}&=&-F_{\hat{\phi}\theta}=B_{\hat{r}(P)}=\frac{1}{2}B\cos{\theta_{s}}\\
    F_{\hat{r}\hat{\phi}}&=&-F_{\hat{\phi}r}=B_{\hat{\theta}(P)}=\frac{1}{2}B\sqrt{f(r_{s})}\sin^2{\theta_{s}}
\end{eqnarray}

In the fluid's rest frame (hereafter, F-frame), the components of the magnetic field take the following form 
\begin{eqnarray}
   B_{\hat{r}(F)}&=& \left(\cos^2{\chi}+\gamma \sin^2{\chi}\right)B_{\hat{r}(P)}\\
   B_{\hat{\theta}(F)}&=&\gamma B_{\hat{\theta}(P)}
\end{eqnarray}
where the Lorentz factor is $\gamma=1/\sqrt{1-\beta^2}$ and the velocity vector is given by
\begin{equation}
    \vec{\beta}=\beta \left(\cos{\chi}\hat{r}+\sin{\chi}\hat{\phi}\right)
\end{equation}
with $\chi$ being the angle between the velocity vector and the radial direction. In our case, where the emitter lies in the equatorial plane at $\theta_s = \pi/2$, the radial component of the magnetic field, $B_{\hat{r}}$, vanishes. Thus, the only non-zero component of the magnetic field is
\begin{equation}
B_{\hat{\theta}(F)} = \frac{\gamma}{2} B \sqrt{f(r_s)},
\end{equation}
which corresponds to the vertical magnetic field in the fluid frame.

Now let us consider polarization vector $\vec{f}_{(F)}$ in the $F$-frame as a vector orthogonal to local magnetic field $\vec{B}_{(F)}=\left(B_{\hat{r}},B_{\hat{\phi}},B_{\hat{\theta}}\right)$ and  wave vector $\vec{k}_{(F)}=\left(k_{\hat{r}},k_{\hat{\phi}},k_{\hat{\theta}}\right)$ direction 
\begin{equation}    \vec{f}_{(F)}=\frac{\left(\vec{k}_{F}\times \vec{B}_{F}\right)}{|\vec{k}_{F}|}, \quad f_{t(F)}=0.
\end{equation}

In this scenario, the intensity of synchrotron radiation along wave vector is proportional to
\begin{equation}   f^{\mu}f_{\mu}=\sin^2{\zeta}|\vec{B}_{(F)}|^2
\end{equation}
with pitch angle
\begin{equation}
   \sin{\zeta}= \frac{|\vec{k}_{(F)}\times \vec{B}_{(F)}|}{|\vec{k}_{(F)}||\vec{B}_{(F)}|}
\end{equation}
One can use inverse Lorentz boost to transform polarization vector which yields
\begin{eqnarray}
    f_{(P)}^{\hat{t}}&=&\gamma f^{\hat{t}}_{(F)}+\gamma \beta \cos{\chi}f^{\hat{r}}_{(F)}+\gamma \beta \sin{\chi}f^{\hat{\phi}}_{(F)}\\
    f_{(P)}^{\hat{r}}&=&\gamma\beta \cos{\chi} f^{\hat{t}}_{(F)}+(1+\left(\gamma-1\right)\cos^2{\chi})f_{(F)}^{\hat{r}}\\
    & &+(\gamma-1)\cos{\chi}\sin{\chi}f_{(F)}^{\hat{\phi}}\nonumber\\
    f_{(P)}^{\hat{\phi}}&=&\gamma\beta \sin{\chi} f^{\hat{t}}_{(F)}+(\gamma-1)\cos{\chi}\sin{\chi}f_{(F)}^{\hat{r}}\\
    & &+(1+\left(\gamma-1\right)\sin^2{\chi})f_{(F)}^{\hat{\phi}}\nonumber\\
    f_{(P)}^{\hat{\theta}}&=&f_{(F)}^{\hat{\theta}}
\end{eqnarray}
In further calculations, it is important to know the polarization vector expressed in global coordinates, which is obtained by applying the inverse transformation from the local $P$-frame.
\begin{eqnarray}
    f^{t}&=&\frac{f^{\hat{t}}_{(P)}}{\sqrt{f(r_{s})}},\quad f^{r}=\sqrt{f(r_{s})}f^{\hat{r}}_{(P)},\nonumber\\
    f^{\phi}&=&\frac{f^{\hat{\phi}}_{(P)}}{r_{s}},\quad f^{\theta}=\frac{f^{\hat{\theta}}_{(P)}}{r_{s}}.
\end{eqnarray}

\subsection{Propagation of linearly polarized light} 
 Light travailing from source to observer undergo redshift which leads following expression for momentum of photon in geodesic local Cartesian $P$-frame
\begin{eqnarray}
k^{\hat{t}}_{(P)} &=& \frac{1}{\sqrt{f(r_s)}}, \quad
k^{\hat{\phi}}_{(P)} = -\frac{\sin\xi \sin\alpha}{\sqrt{f(r_s)}}, \nonumber \\
k^{\hat{r}}_{(P)} &=& \frac{\cos\alpha}{\sqrt{f(r_s)}}, \quad
k^{\hat{\theta}}_{(P)} = \frac{\cos\xi \sin\alpha}{\sqrt{f(r_s)}}.
\end{eqnarray}
Here $\alpha$ is the emission angle, and $\xi$ is defined by
\begin{equation}
\cos\xi = \frac{\cos\theta_0}{\sin\psi}, \qquad
\sin\xi = \frac{\sin\theta_0 \cos\phi}{\sin\psi}.
\end{equation}
where $\psi-$ angle between $\hat{r}$ and unit vector $\hat{n}$ towards the observer.

At the next step, one must rewrite the emission angle $\alpha$ in terms of $\psi$. To do so, we calculate the angle $\psi$ by assuming the observer is located at infinity as
\begin{equation}    \psi=\int_{r_{s}}^{\infty}\frac{\rm{d} r}{r^2\sqrt{\frac{1}{b^2}-\frac{f(r)}{r^2}}}\ ,
\end{equation}
together with
\begin{equation}
    \sin{\alpha}=\frac{b}{r_{s}}\sqrt{f(r_{s})} .
    \label{sin_rel}
\end{equation}
which helps us to find the relationship between $\psi$ and $\alpha$ emission angle. The relation \eqref{sin_rel} is found by using the fact $\tan{\alpha}=\sqrt{u^{\psi}u_{\psi}}/\sqrt{u^{r}u_{r}}$ \citep[for more details see][]{2002ApJ...566L..85B}. Finally, we obtain the 4-momentum of the photon in the fluid frame by using the Lorentz transformations as
\begin{eqnarray}
    k_{(F)}^{\hat{t}}&=&\gamma k_{P}^{\hat{t}}-\gamma \beta \cos{\chi}k^{\hat{r}}_{P}-\gamma \beta \sin{\chi}k^{\hat{\phi}}_{(P)} \, ,\\
     k_{(F)}^{\hat{r}}&=&-\gamma \beta \cos{\chi}k^{\hat{t}}_{P}+\left(1+(\gamma-1)\cos^2{\chi}\right)k^{\hat{r}}_{(P)}\\
     & &+(\gamma-1)\cos{\chi}\sin{\chi}k_{(P)}^{\hat{\phi}} \, , \nonumber\\
     k_{(F)}^{\hat{\phi}}&=&-\gamma \beta \sin{\chi}k^{\hat{t}}_{P}+(\gamma-1)\cos{\chi}\sin{\chi}k_{(P)}^{\hat{r}}\\
     & &+\left(1+(\gamma-1)\cos^2{\chi}\right)k^{\hat{\phi}}_{(P)} \, , \nonumber\\
     k_{(F)}^{\hat{\theta}}&=&k_{(P)}^{\hat{\theta}} \, .
\end{eqnarray}
\subsection{Stokes parameters}
We are now ready to introduce the complex Penrose-Walker constant \citep{1970CMaPh..18..265W}

\begin{eqnarray}  
    \kappa&=&\kappa_{1}+i\kappa_{2},\\
    \kappa_{1}&=&\psi_{2}^{-1/3}\left(k^{t}f^{r}-k^{r}f^{t}\right),\\ \kappa_{2}&=&\psi_{2}^{-1/3}r_{s}^2\left(k^{\phi}f^{\theta}-k^{\theta}f^{\phi}\right),\\\nonumber
\end{eqnarray}

where the photon momentum and polarization components are expressed in Boyer-Lindquist coordinates. These components are obtained via an inverse transformation from the local frame:
\begin{eqnarray}
    k^{t}&=&\frac{k^{\hat{t}}_{(P)}}{\sqrt{f(r_{s})}},\quad k^{r}=\sqrt{f(r_{s})}k^{\hat{r}}_{(P)},\nonumber\\
    k^{\phi}&=&\frac{k^{\hat{\phi}}_{(P)}}{r_{s}},\quad k^{\theta}=\frac{k^{\hat{\theta}}_{(P)}}{r_{s}} \, .
\end{eqnarray}

together with Weyl scalar

\begin{equation}
    \psi_{2}=\frac{M}{r_{s}^{3}}.
\end{equation}

The celestial coordinates $(x,y)$ for a photon moving towards the observer at infinity are described as follows:

\begin{eqnarray}
    x&=&-\frac{r_{s}k^{\hat{\phi}}_{(P)}}{\sin{\theta}},\\
    y&=&r_{s}\sqrt{\left(k^{\hat{\theta}}_{(P)}\right)^2-\left(k^{\hat{\phi}}_{(P)}\right)^2\cot^2\theta}\, \operatorname{sgn} 
    \left(\sin{\phi}\right).
\end{eqnarray}

The expression for the normalized electric field $\vec{E}$ associated with synchrotron radiation, as it propagates along a geodesic in the observer's sky coordinates, is given as follows:

\begin{eqnarray}
    E_{x,\mathrm{norm}}&=&\frac{y\kappa_{2}+x\kappa_{1}}{\sqrt{\left(\kappa_{1}^2+\kappa_{2}^2\right)\left(x^2+y^2\right)}} , \\
    E_{y,\mathrm{norm}}&=&\frac{y\kappa_{1}-x\kappa_{2}}{\sqrt{\left(\kappa_{1}^2+\kappa_{2}^2\right)\left(x^2+y^2\right)}} , \\
    E_{x,\mathrm{norm}}^2&+&E_{y,\mathrm{norm}}^2=1.
\end{eqnarray}

\begin{figure*}[t]
\centering
\includegraphics[width=0.3\textwidth]{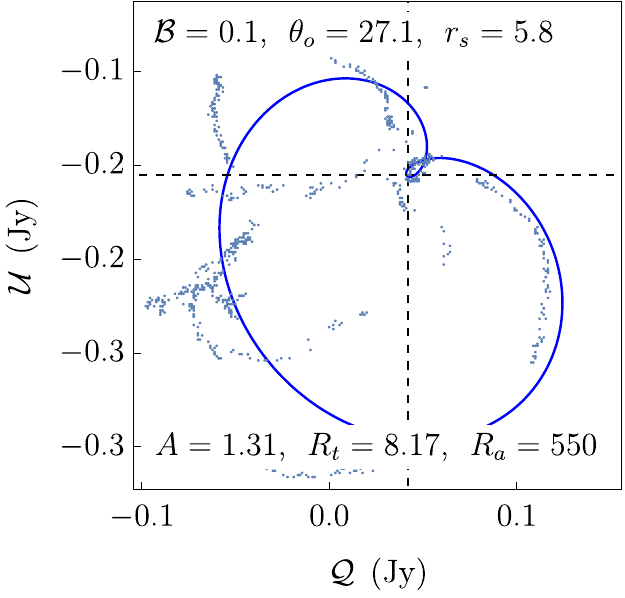}
\includegraphics[width=0.3\textwidth]{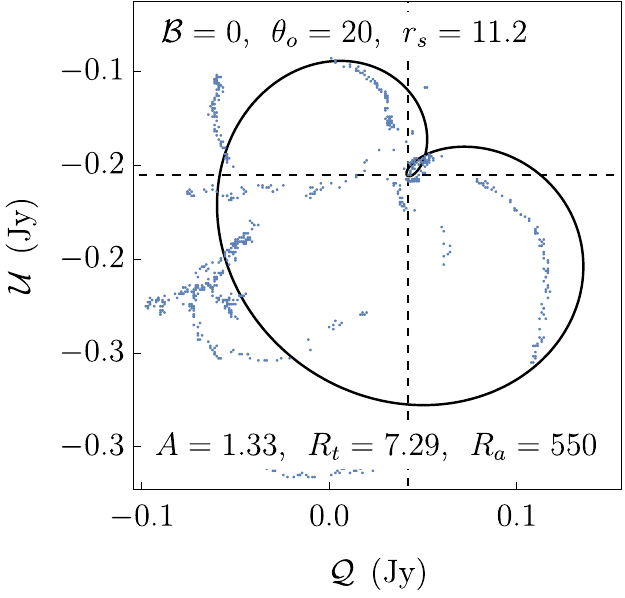}
\includegraphics[width=0.3\textwidth]{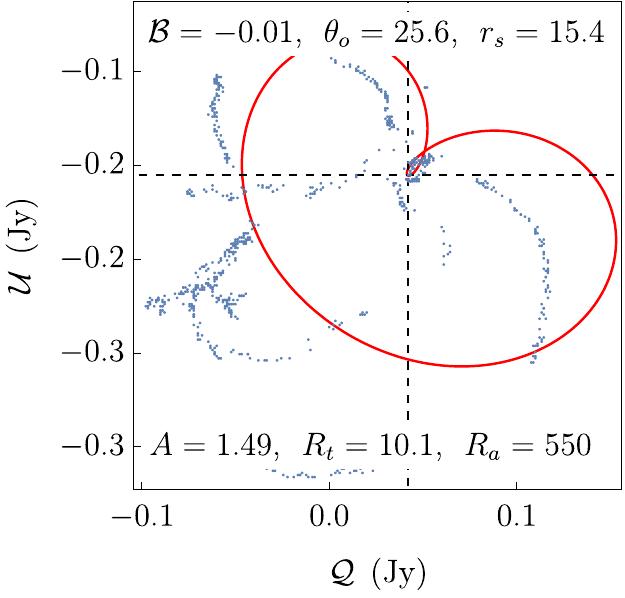}
\caption{Comparison between ALMA $Q-U$ loop data with semi-analytical ring model prediction for $\BB =0.1$ (left), $\BB =0$ (middle), and $\BB =-0.01$ (right). 
}
\label{Q_U_pic}
\end{figure*} 

The components of the observed electric field and the intensity of synchrotron radiation, as received by an observer at infinity, are dependent on multiple factors and are expressed as:

\begin{eqnarray}    
    E_{x,\mathrm{obs}}&=&\delta^{\frac{3+\alpha_{\nu}}{2}}l_{p}^{1/2}|\vec{B}|^{\frac{1+\alpha_{\nu}}{2}}\sin^{\frac{1+\alpha_{\nu}}{2}}{\zeta}E_{x,\mathrm{norm}},\\
    E_{y,\mathrm{obs}}&=&\delta^{\frac{3+\alpha_{\nu}}{2}}l_{p}^{1/2}|\vec{B}|^{\frac{1+\alpha_{\nu}}{2}}\sin^{\frac{1+\alpha_{\nu}}{2}}{\zeta}E_{y,\mathrm{norm}},\\
    E_{x,\mathrm{obs}}^2&+&E_{y,\mathrm{obs}}^2= \delta^{3+\alpha_{\nu}}l_{p}|\vec{B}|^{1+\alpha_{\nu}}\sin^{1+\alpha_{\nu}}{\zeta},
\end{eqnarray}

where $\delta=1/k_{(\rm{F})}^{\hat{t}}$ is the Doppler shift factor and $\alpha_{\nu}$ is the spectral index. 
In the following, we assume the emitted spectrum follows a power-law distribution $I_{\nu} \propto \nu^{-\alpha_{\nu}}$. As a consequence, the angular dependence of the emission varies as $\sin^{1+\alpha_{\nu}}\zeta$. Therefore, for spectral index $\alpha_{\nu}$, 
the observed flux density is related to the hot-spot's emitted flux density by
\begin{equation}
\frac{I_{\nu,\mathrm{obs}}}{I_{\nu,\mathrm{HS}}} = \delta^{3+\alpha_{\nu}}.
\label{Iobs}
\end{equation} 
The other quantity $l_{p}$ describes photon path length, which is usually written in the form
\begin{equation}   
    l_{p}=\frac{k^{\hat{t}}_{(F)}}{k^{\hat{\theta}}_{(F)}}H,
\end{equation}
where $H$ is the height of the disk. In the following, we set $\alpha_{\nu}=1$ a representative value at $230\,\rm{GHz}$ frequency \citep{2021ApJ...910L..14G}. As a result, the intensity $I$ and EVPA of polarized light becomes  
\begin{eqnarray}    
\label{I_EVPA_eq}
    I&=&|P|=\delta^{4}l_{p}|\vec{B}|^2\sin^2{\zeta},\\
    EVPA&=&0.5\mathrm{Arg}(P)=0.5\mathrm{Arg}{\left(Q+i U\right)}\, , 
\end{eqnarray}
where the Stokes parameters $Q$ and $U$ are defined as
\begin{equation}
    Q=E_{y,\mathrm{obs}}^2-E_{x,\mathrm{obs}}^2\ , \quad U=-2E_{x,\mathrm{obs}}E_{y,\mathrm{obs}}.
    \label{stokes_par}
\end{equation}
In our model, circular polarization is neglected. 
Moreover, the linear polarization fraction is equal to a constant $p$ and total intensity of radiation satisfies $\sqrt{Q^2+U^2}/I=p$. We take $p=1$ for the sake of simplicity, hence $|P| = \sqrt{Q^2 + U^2} = I$.  The Stokes parameters $Q$ and $U$ are calculated using equation \eqref{stokes_par}. 

\section{Results for an orbiting point-like hot spot}\label{sec:parameters}

For hot-spot modeling, we assume a localized isotropic emitting region of characteristic size $L \sim r_g$, confined to the equatorial plane. The primary radiative transfer and polarization calculations are performed in the point-source approximation, in which only the trajectory of the centroid of the emitting region is followed. This approximation is justified provided that the source size remains small compared to the characteristic spatial scales over which relevant quantities vary along the orbit, including the magnetic-field structure, gravitational lensing, and relativistic polarization transport across neighboring geodesics.

Effects of extended size are incorporated in the depolarization calculation by integrating over the spatial extent of the equatorial emitting patch. These effects become significant when relevant quantities vary appreciably across the emitting region. We address this in Sec.~\ref{sec:depol}.

\subsection{Effect of electromagnetic interaction on polarization}

We are now ready to construct the polarization images of the hot-spot orbiting black hole in the equatorial plane, at a fixed radius $r=r_{s}$. We choose the motion to be the clockwise, corresponding to an angle $\chi=-\pi/2$. The associated relativistic velocity measured in the locally non-rotating frame is given by 
\begin{equation}
    \beta_{R}=\Omega\frac{r_{s}}{\sqrt{f(r_{s})}},
\end{equation}
where angular velocity $\Omega$ is defined by the formula \eqref{ang_vel} in the previous section. 
The period of the hot-spot orbiting black hole can be determined using the relation 
\begin{equation}
    T=2\pi t_{g}\sqrt{r^3\left(2\mathcal{B}^2r^2(r-2)+2\mathcal{B}rF+1\right)} \, ,
\end{equation} 
where $t_{g}=GM/c^3$ is the gravitational timescale and $\BB$ is a magnetic field interaction parameter defined in (\ref{eq:param}). 
\begin{figure*}
\centering
\includegraphics[width=0.9\textwidth]{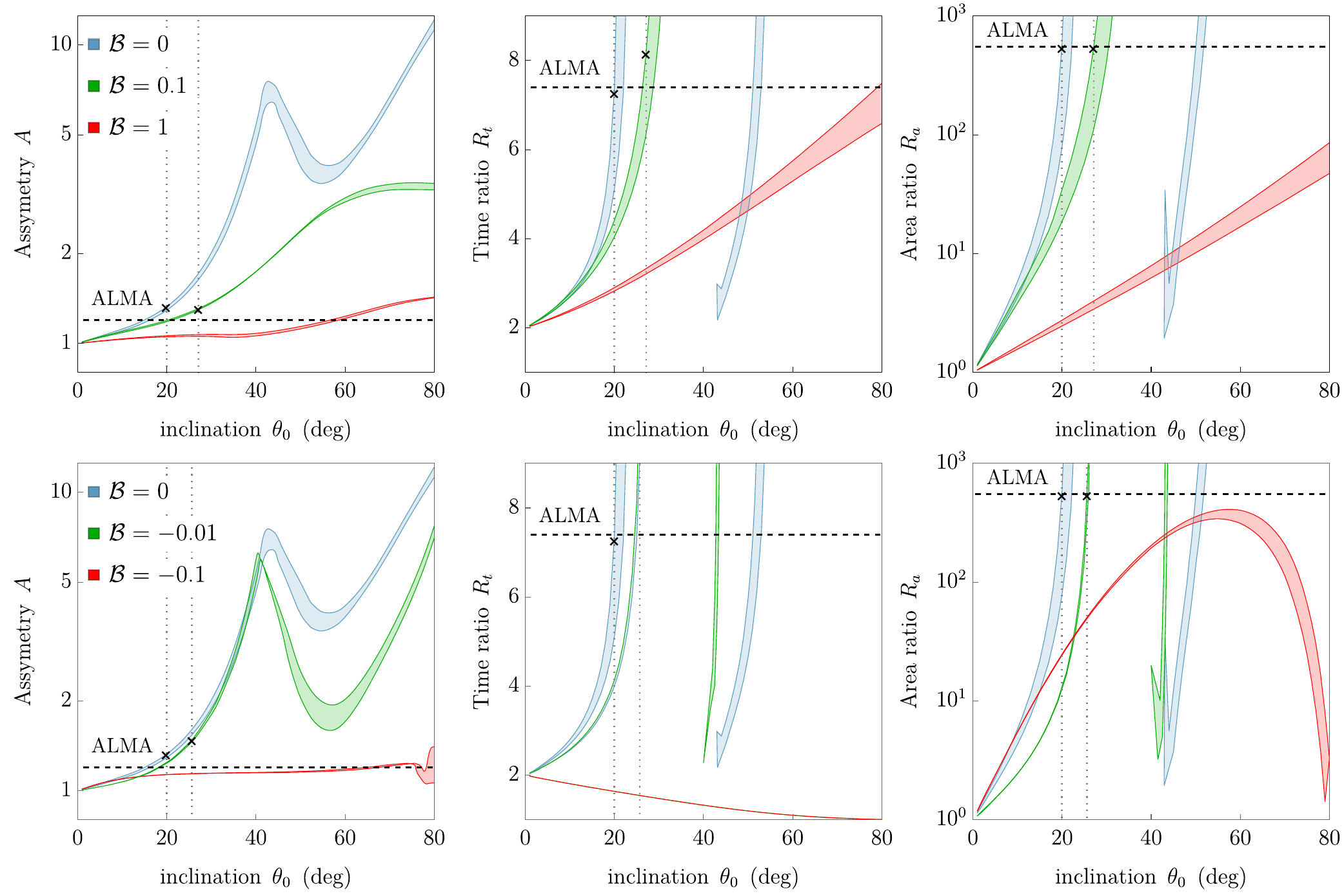}
\caption{Asymmetry $A$, time $R_{t}$ and area $R_{a}$ ratio in case of negative magnetic parameters (top) and positive (bottom) with respect to inclination angle. The smallest inclination angle at which $R_{a} = 550$ is indicated by a cross.}
\label{fig_constrain}
\end{figure*}
In Figures~\ref{polarimatic_image_pic} and \ref{polarimatic_image_pic2} we illustrate how the magnetic interaction parameter $\BB$ affects every major polarization observable for an orbiting hot-spot, and how those effects vary with inclination angle. 
From left to right we show the instantaneous polarization vector maps on the sky (the ticks are normalized to the same maximum length in each panel), the total polarization intensity, the EVPA, and the $Q$–$U$ trajectory over one full $2\pi$ of azimuthal phase $\phi$. 
Rows correspond to the inclinations of $20^{\circ}$, 
$40^{\circ}$, 
$60^{\circ}$  (from top to bottom), while colors show 
$\BB=+0.01$ (blue), 
$0$ (black), $-0.01$ (red), that is the interaction parameter is relatively weak. Fig.~\ref{polarimatic_image_pic} show the hot-spot at radius $r=11\,r_{g}$, while Fig. \ref{polarimatic_image_pic2} repeats the same analysis at $r = 6\,r_{g}$, bringing the hot‐spot deeper into the strong‐gravity zone. 

In the left panels of Figures~\ref{polarimatic_image_pic} and \ref{polarimatic_image_pic2} illustrating the polarization patterns, systematic offsets (zoomed) are visible: a positive $\BB$ (blue) twists the polarization direction forward in the direction of motion, while negative $\BB$ (red) twists it backward. Changing the viewing angle shows that at low inclinations the local magnetic interaction torque more directly alters the projected field geometry, whereas at high inclinations the lensing effect dominates. The total intensity $I$ plots exhibit peaks when the emitting hot-spot passes behind the black hole due to maximal Doppler boosting. As with the other observables, the absolute difference between positive and negative $\BB$ grows as the system is viewed more face-on. This shows that low‐inclination polarized flares provide a better sensitivity to the magnetic interaction parameter $\BB$. 

The third columns of Figures~\ref{polarimatic_image_pic} and \ref{polarimatic_image_pic2} shows the EVPA versus orbital phase. Interestingly, even small $\BB = \pm 0.01$ shifts the entire EVPA curve $\sim 5$–$10^\circ$ and introduces a slight asymmetry between the incoming and outgoing rotations. Since the EVPA can in principle be measured with sub-degree precision in bright flares, even a small persistent offset can potentially provide a sensitive probe of $\BB$. A limiting factor at  230 GHz frequency is the variable internal Faraday screen of Sgr~A* \citep{2024A&A...682A..97W}, which we neglect in this analysis. Finally, the rightmost panels of Figures~\ref{polarimatic_image_pic} and \ref{polarimatic_image_pic2} illustrate the $Q$–$U$ trajectories over one orbit. The magnetic interaction parameter does not change the sense of the loop rotation, but its shape and enclosed area clearly depend on the sign and magnitude of $\BB$. We quantify this in the next section when applied to the observational data.

\subsection{Polarization loops from Sgr~A* with ALMA} 

\cite{2022A&A...665L...6W} reported an exceptional radio flare from Sgr~A* detected by the ALMA, whose EVPA traces clear, quasi‐closed loops in the Stokes $Q$-$U$ plane.  By fitting a synchrotron hot‐spot orbiting a black hole, they extracted three key parameters of each loop: asymmetry $A$ -- the largest ratio of perpendicularly projected diameters of the loops, time ratio $R_{t}$ -- the ratio of the full loop period to the time, and area ratio $R_{a}$ -- the ratio of area of the large loop to that of the small loop. Based on the ALMA measurements \cite{2022A&A...665L...6W} constrained these parameters to  
\begin{equation}
  A = 1.3 \pm 0.1,\quad
  R_{t} = 6.5 \pm 0.5,\quad
  R_{a} = 550 \pm 50,
\end{equation} 
and inferred an observer inclination $\theta_0\simeq20^\circ$. They also derived that the magnetic field near black hole is dominated by the vertical component. 

We extend their analysis by incorporating electromagnetic interaction between the synchrotron-emitting hot-spot and those vertical magnetic fields parametrized by $\BB$. If one considers $\mathcal{B}=0$, corresponding to a purely Keplerian hot-spot, the emitter is located at approximately $r\approx 10.6\pm0.6 \,r_{g}$, corresponding to a range from $r_{\rm{min}}=10r_{g}$ to $r_{\rm{max}}=11.2r_{g}$.
However, for nonzero $\mathcal{B}$, the situation changes considerably, as we demonstrate in Figures~\ref{Q_U_pic} and \ref{fig_constrain}.

Figure \ref{Q_U_pic} shows a comparison between the $Q$-$U$ fluxes predicted by the simple ring model and the ALMA observations acquired on April 11, 2017 for different $\BB$. In Figure \ref{fig_constrain} we show the dependence of $A$, $R_{t}$, $R_{a}$ for different values of the inclination angle. The overall result is consistent with that obtained by \cite{2022A&A...665L...6W}, and it predicts a small tilt angle of approximately $\theta_{o}\approx20^{\circ}$ for $\mathcal{B}=0$,  $\theta_{\circ}\approx26^{\circ}$ for $\mathcal{B}=-0.01$ and $\theta_{\circ}\approx27^{\circ}$ for $\mathcal{B}=0.1$, respectively. Interestingly, including either the positive or negative magnetic parameter, both leads to a larger inclination angles. In our comparison, it is evident that large negative values of the magnetic parameter around $\mathcal{B}\approx-0.1$ fail to accurately reproduce the time ratio observed in the ALMA data (see bottom middle panel). Likewise, large positive values of $\mathcal{B}$ also struggle to reproduce the area ratio (see top right panel). Therefore, one can infer that the magnetic parameter likely lies within the range $-0.1<\mathcal{B}<1$.

In the figures, the area ratio of the loops $R_{a}=550$ remains unchanged. In this scenario, in order to better reproduce the observations, we set $r_{\rm{s}}=r_{\rm{max}}$ in each figure, corresponding to the smallest tilt angle (as illustrated in Figure \ref{fig_constrain}, smaller $\theta_{o}$ produce more symmetric loops). As shown in the figures, 
introducing a positive magnetic parameter makes the loop more symmetrical, although it also increases the $R_{t}$ ratio. In the case of a negative magnetic parameter, the situation changes dramatically. Including a small negative $\mathcal{B}$ increases not only the time ratio but also makes the loops more asymmetric. Therefore, it can be inferred that if a magnetic parameter is present, it likely has a positive effective sign, corresponding to either $q>0$, $B>0$ or $q<0$, $B<0$. 

\section{Depolarization of a finite-size hot spot} \label{sec:depol}
We quantify the depolarization arising from the non-zero spatial extent of a synchrotron-emitting hot spot in the equatorial plane of a spherical symmetric spacetime by comparing its net linear polarization to that of an equivalent point-like emitter. We consider a symmetric, optically thin hot spot centered at radius $r_c$ and azimuth $\phi_c=-\Omega_{c}t$ at emission time $t$, with emissivity profile scaling modeled as 
\begin{align}
J(r,\phi) \sim \exp\left[-4\log{2}\left(\frac{ \Delta x^2+\Delta y^2}{w^2}\right)\right]=\nonumber\\
\exp\left[-4\log{2}\left(\frac{\Delta r^2+4r_{c}\left(r_{c}+\Delta r\right)\sin^2{\frac{\Delta \phi}{2}}}{w^2}\right)\right],
\end{align}
where the minus sign in $\phi_c = -\Omega_c t$ denotes clockwise motion, $\Delta\phi = \arg{e^{i(\phi-\phi_c)}}$ is the wrapped phase difference, and $\Delta r = r - r_c$ is the radial displacement from the spot center. The parameter $w$ is the full width at half maximum (FWHM) of the Gaussian profile. In the following, we identify the hot-spot scale as $L \sim w$.

The observed Stokes parameters are computed by incorporating all relevant general relativistic radiative transfer effects. The observed intensity on the image plane is expressed as
\begin{equation}
\label{I_{obs}}
I(\rho,\varphi)=J(\rho,\varphi)I_{\nu,\mathrm{obs}},
\end{equation}
\begin{figure}[t]
\centering
\includegraphics[width=0.8\linewidth]{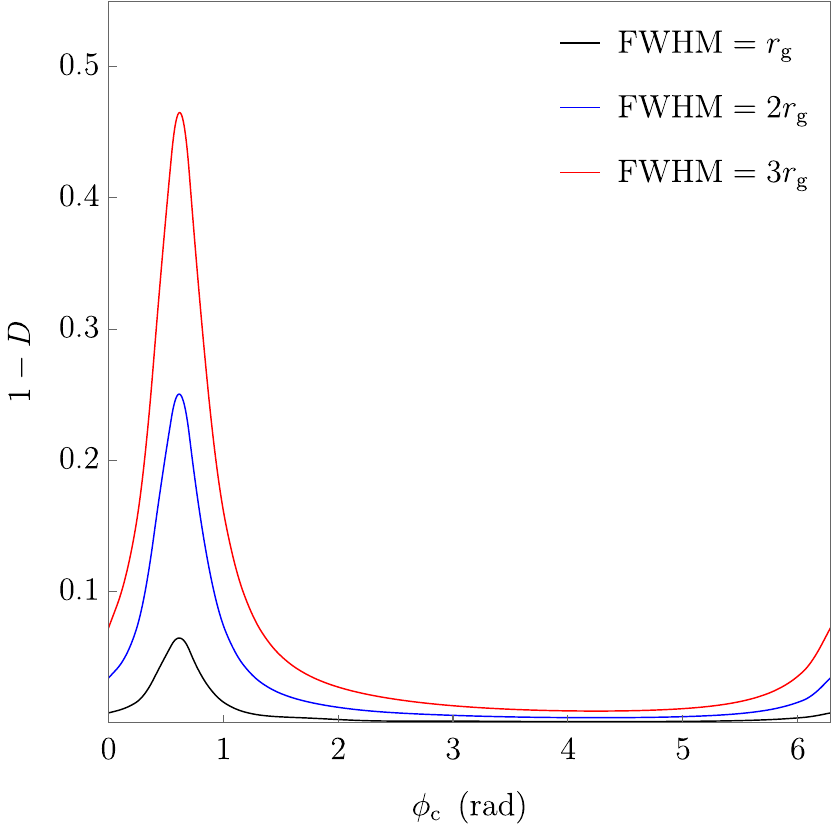}
\caption{
Depolarization of the emitted radiation as a function of the orbital phase $\phi_{c}$ for an orbiting hot spot with different size measured with FWHM. The observer inclination angle is fixed at $\theta_{o}=20^\circ$, while the hot-spot center is located at $r_{\rm c}=10~r_{g}$. The figure demonstrates how the extent of the emitting region influences the phase-dependent depolarization behavior.} 
\label{dipol_pic}
\end{figure} 
where $(\rho,\varphi)$ denote the polar coordinates on the observer's screen, defined as
\begin{equation}
\rho=\sqrt{x^{2}+y^{2}}, \qquad
\varphi = \arg(x + iy).
\end{equation}
In evaluating Eqs.~\eqref{I_EVPA_eq}--\eqref{stokes_par}, we include the redshift correction through the observed specific intensity $I_{\nu,\mathrm{obs}}$ defined by Eq.~\eqref{Iobs}.

The net observed polarization from the extended hot spot is obtained by integrating the Stokes parameters over the image of the emitting region,
\begin{align}
&Q_{\mathrm{tot}} = \iint Q(\rho,\varphi) dA, \quad
U_{\mathrm{tot}} = \iint U(\rho,\varphi) dA, \\
&I_{\mathrm{tot}} = \iint I(\rho,\varphi) dA,
\end{align}
leading to a polarization fraction
\begin{equation}
p_{\mathrm{finite}} = \frac{\left| Q_{\rm tot} + i\,U_{\rm tot} \right| }{I_{\rm tot}} = 
\frac{\sqrt{Q_{\mathrm{tot}}^2 + U_{\mathrm{tot}}^2}}{I_{\mathrm{tot}}}.
\end{equation}
For comparison, the polarization fraction of a point-like emitter located at $(r_c,\phi_{c})$ is $p_{\mathrm{point}}=1$. Therefore, we define the depolarization factor as
\begin{equation}
D = \frac{p_{\mathrm{finite}}}{p_{\mathrm{point}}}.
\end{equation}
 
The net observed Stokes parameters are obtained by numerically integrating over the observer’s image plane. In practice, we sum local flux densities $I(x,y)$, $Q(x,y)$, and $U(x,y)$ over all pixels in an image with 30$\,r_g$ field of view and resolution $512 \times 512$. Depolarization arises from spatial variations of the EVPA across the image of the emitting region, leading to partial cancellation in the vector sum of local polarized fluxes. For \(w = r_g\) and \(r_c = 10\,r_g\), the variation of the EVPA across the hot spot is small, resulting in weak depolarization. The net depolarization fraction is given by
\begin{eqnarray}
    (1-\langle D \rangle)\times100\% \approx 0.72\% ,
\end{eqnarray}
where \(\langle D \rangle\) is the time-averaged depolarization factor over one orbit, which remains close to unity.

In Fig.~\ref{dipol_pic}, the weakest finite-size depolarization occurs near $\phi_{c}\sim 7\pi/5$, corresponding to the hot-spot position in front of the black hole. In contrast, the depolarization becomes strongest around $\phi_{c}\sim\pi/5$, when the hot spot is located behind the black hole. In this configuration, photons emitted from different regions of the source propagate along substantially different geodesics due to strong gravitational lensing, enhancing variations in the polarization properties across the emitting region and leading to a stronger net depolarization effect. For comparison, in the case of a static hot spot the depolarization extrema occur exactly at $\phi_{c}=\pi/2$ (maximum) and $\phi_{c}=3\pi/2$ (minimum). The slight shift of these extrema in the orbiting case is caused by the nonzero orbital velocity of the emitting plasma. We note, however, that this effect becomes significant for emission originating close to the black hole, where strong-field lensing amplifies variations in the transported polarization vectors. Our results indicate that strong depolarization occurs when the characteristic source size becomes comparable to the distance of the source center from the black hole, namely $w/r_c \sim 1$. In this regime, differential transport across the emitting region becomes substantial, producing large variations in the transported polarization directions and consequently leading to a significant reduction of the integrated polarization degree.

\section{Conclusions}

Our results demonstrate that even within a quasi-neutral plasma, small charge imbalances can lead to electromagnetic interactions with significant dynamical and observational consequences near Sgr~A*. In particular, the new magnetic interaction parameter $\BB$ produces characteristic, inclination‐dependent imprints on all polarization observables of an orbiting hot-spot. In practice, the strongest effects can be measurable from modeling the EVPA offset and the $Q$–$U$ loop area.

In this study, we adopt two distinct simplifications: (1) the black hole is assumed to be non-rotating, and (2) only the primary (direct) image is considered, while higher-order images are neglected. However, black hole spin can contribute noticeably, and secondary images may influence the detailed structure of the $Q$–$U$ loop \citep{2021PhRvD.104d4060G,2024A&A...685A.142Y}. Including them would help capture spin-dependent effects more accurately, but treating higher-order images is
challenging within this framework, so they are not included in this analysis. Moreover, in a rotating Kerr black hole, the frame-dragging of spacetime twists the magnetic field lines, inducing an electric field. This effect is encoded in Wald’s solution as a nonzero temporal component of the vector potential, which gives rise to a radial and polar electric field near the black hole \citep{2022MNRAS.512.2798K}. 

Our results indicate that the depolarization associated with the finite extent of the emitting region is weak for compact hot spots, with the orbit-averaged value remaining below $1\%$ and maximum along the orbit of around 5\% for $w \sim r_g$ at orbital radius $r_c = 10\,r_g$. However, a systematic increase is observed for larger hot-spot sizes ($w > r_g$), leading to more pronounced depolarization, reaching orbit-averaged value of $\approx 6.4\%$ and a maximum along the orbit of over 40\% for $w = 3\,r_g$.

Inclusion of the $\BB$‐term in our emission model represents a direct coupling between the hot‐spot electrons and an external, large‐scale magnetic field around the black hole. A nonzero $\BB$ applies a small torque on the emitting electrons, tilting their pitch‐angle distribution and hence the instantaneous polarization vector.  As the hot-spot orbits, this torque accumulates a phase shift that appears as a coherent EVPA offset and altered $Q$–$U$ morphology. Simultaneously the model allows the emission to occur closer to the black hole.  

Even a small charge imbalance in the hot-spot (e.g., due to charge separation originating from turbulence) renders it sensitive to a Lorentz force that can subtly alter its trajectory, and hence its EVPA features. We conclude that only weak electromagnetic coupling, $|\mathcal{B}|\lesssim0.1$ remains consistent with ALMA’s polarization loops. These limits on $\mathcal{B}$ translate into constraints on the product of the hot‐spot charge‐to‐mass ratio and the ambient field strength at radii $r\sim5$--$10\,r_{g}$. In the absence of electromagnetic interaction, stable circular orbits around a Schwarzschild black hole are only possible at radii greater than $r=6\,r_{g}$. Therefore, any emission originating from $r<6\,r_{g}$ may result from electromagnetic effects, which shift the location of stable orbits closer to the event horizon. Future high-precision astrometric monitoring will directly localize the hot-spot’s orbital phase and radius, breaking the current degeneracy with inclination and $\BB$ and thereby tightening constraints on both the coupling parameter and the magnetic field strength.

\begin{acknowledgements} 
We thank the anonymous referee for the valuable comments and suggestions, which helped improve the quality of the manuscript. 
A. Tlemissov and A. Tursunov acknowledge support from the internal grants of the Silesian University in Opava No. IGS/27/2026 and SGS/24/2024. 
A. Tlemissov also acknowledges  the support of science and research in the Moravian-Silesian Region 
Grant No.  RRC/09/2023. MW is supported by a~Ramón y Cajal grant RYC2023-042988-I from the Spanish Ministry of Science and Innovation and acknowledges financial support from the Severo Ochoa grant CEX2021-001131-S funded by MCIN/AEI/ 10.13039/501100011033.
\end{acknowledgements}



\begin{thebibliography}{42}
    \expandafter\ifx\csname natexlab\endcsname\relax\def\natexlab#1{#1}\fi
    
    \bibitem[{{Beloborodov}(2002)}]{2002ApJ...566L..85B}
    {Beloborodov}, A.~M. 2002, \apjl, 566, L85
    
    \bibitem[{{Blandford} \& {Znajek}(1977)}]{1977MNRAS.179..433B}
    {Blandford}, R.~D. \& {Znajek}, R.~L. 1977, \mnras, 179, 433
    
    \bibitem[{{Chen}(2016)}]{2016ippc.book.....C}
    {Chen}, F.~F. 2016, {Introduction to Plasma Physics and Controlled Fusion}
    
    \bibitem[{{Eckart} {et~al.}(2012){Eckart}, {Garc{\'{\i}}a-Mar{\'{\i}}n}, {Vogel}, {Teuben}, {Morris}, {Baganoff}, {Dexter}, {Sch{\"o}del}, {Witzel}, {Valencia-S.}, {Karas}, {Kunneriath}, {Straubmeier}, {Moser}, {Sabha}, {Buchholz}, {Zamaninasab}, {Mu{\v z}i{\'c}}, {Moultaka}, \& {Zensus}}]{2012A&A...537A..52E}
    {Eckart}, A., {Garc{\'{\i}}a-Mar{\'{\i}}n}, M., {Vogel}, S.~N., {et~al.} 2012, \aap, 537, A52
    
    \bibitem[{{Event Horizon Telescope Collaboration} {et~al.}(2024{\natexlab{a}}){Event Horizon Telescope Collaboration}, {Akiyama}, {Alberdi}, {Alef}, {Algaba}, {Anantua}, {Asada}, {Azulay}, {Bach}, {Baczko}, {Ball}, {Balokovic}, {Bandyopadhyay}, {Barrett}, {Baub{\"o}ck}, {Benson}, {Bintley}, {Blackburn}, {Blundell}, {Bouman}, {Bower}, {Boyce}, {Bremer}, {Brinkerink}, {Brissenden}, {Britzen}, {Broderick}, {Broguiere}, {Bronzwaer}, {Bustamante}, {Byun}, {Carlstrom}, {Ceccobello}, {Chael}, {Chan}, {Chang}, {Chatterjee}, {Chatterjee}, {Chen}, {Chen}, {Cheng}, {Cho}, {Christian}, {Conroy}, {Conway}, {Cordes}, {Crawford}, {Crew}, {Cruz-Osorio}, {Cui}, {Dahale}, {Davelaar}, {De Laurentis}, {Deane}, {Dempsey}, {Desvignes}, {Dexter}, {Dhruv}, {Dihingia}, {Doeleman}, {Dougal}, {Dzib}, {Eatough}, {Emami}, {Falcke}, {Farah}, {Fish}, {Fomalont}, {Ford}, {Foschi}, {Fraga-Encinas}, {Freeman}, {Friberg}, {Fromm}, {Fuentes}, {Galison}, {Gammie}, {Garc{\'\i}a}, {Gentaz}, {Georgiev}, {Goddi}, {Gold}, {G{\'o}mez-Ruiz},
      {G{\'o}mez}, {Gu}, {Gurwell}, {Hada}, {Haggard}, {Haworth}, {Hecht}, {Hesper}, {Heumann}, {Ho}, {Ho}, {Honma}, {Huang}, {Huang}, {Hughes}, {Ikeda}, {Impellizzeri}, {Inoue}, {Issaoun}, {James}, {Jannuzi}, {Janssen}, {Jeter}, {Jiang}, {Jim{\'e}nez-Rosales}, {Johnson}, {Jorstad}, {Joshi}, {Jung}, {Karami}, {Karuppusamy}, {Kawashima}, {Keating}, {Kettenis}, {Kim}, {Kim}, {Kim}, {Kim}, {Kino}, {Koay}, {Kocherlakota}, {Kofuji}, {Koch}, {Koyama}, {Kramer}, {Kramer}, {Kramer}, {Krichbaum}, {Kuo}, {La Bella}, {Lauer}, {Lee}, {Lee}, {Leung}, {Levis}, {Li}, {Lico}, {Lindahl}, {Lindqvist}, {Lisakov}, {Liu}, {Liu}, {Liuzzo}, {Lo}, {Lobanov}, {Loinard}, {Lonsdale}, {Lowitz}, {Lu}, {MacDonald}, {Mao}, {Marchili}, {Markoff}, {Marrone}, {Marscher}, {Mart{\'\i}-Vidal}, {Matsushita}, {Matthews}, {Medeiros}, {Menten}, {Michalik}, {Mizuno}, {Mizuno}, {Moran}, {Moriyama}, {Moscibrodzka}, {Mulaudzi}, {M{\"u}ller}, {M{\"u}ller}, {Mus}, {Musoke}, {Myserlis}, {Nadolski}, {Nagai}, {Nagar}, {Nakamura}, {Narayanan}, {Natarajan},
      {Nathanail}, {Fuentes}, {Neilsen}, {Neri}, {Ni}, {Noutsos}, {Nowak}, {Oh}, {Okino}, {Olivares}, {Ortiz-Le{\'o}n}, {Oyama}, {{\"O}zel}, {Palumbo}, {Paraschos}, {Park}, {Parsons}, {Patel}, \& {Pen}}]{2024ApJ...964L..25E}
    {Event Horizon Telescope Collaboration}, {Akiyama}, K., {Alberdi}, A., {et~al.} 2024{\natexlab{a}}, \apjl, 964, L25
    
    \bibitem[{{Event Horizon Telescope Collaboration} {et~al.}(2024{\natexlab{b}}){Event Horizon Telescope Collaboration}, {Akiyama}, {Alberdi}, {Alef}, {Algaba}, {Anantua}, {Asada}, {Azulay}, {Bach}, {Baczko}, {Ball}, {Balokovi{\'c}}, {Bandyopadhyay}, {Barrett}, {Baub{\"o}ck}, {Benson}, {Bintley}, {Blackburn}, {Blundell}, {Bouman}, {Bower}, {Boyce}, {Bremer}, {Brinkerink}, {Brissenden}, {Britzen}, {Broderick}, {Broguiere}, {Bronzwaer}, {Bustamante}, {Byun}, {Carlstrom}, {Ceccobello}, {Chael}, {Chan}, {Chang}, {Chatterjee}, {Chatterjee}, {Chen}, {Chen}, {Cheng}, {Cho}, {Christian}, {Conroy}, {Conway}, {Cordes}, {Crawford}, {Crew}, {Cruz-Osorio}, {Cui}, {Dahale}, {Davelaar}, {De Laurentis}, {Deane}, {Dempsey}, {Desvignes}, {Dexter}, {Dhruv}, {Dihingia}, {Doeleman}, {Dougall}, {Dzib}, {Eatough}, {Emami}, {Falcke}, {Farah}, {Fish}, {Fomalont}, {Ford}, {Foschi}, {Fraga-Encinas}, {Freeman}, {Friberg}, {Fromm}, {Fuentes}, {Galison}, {Gammie}, {Garc{\'\i}a}, {Gentaz}, {Georgiev}, {Goddi}, {Gold}, {G{\'o}mez-Ruiz},
      {G{\'o}mez}, {Gu}, {Gurwell}, {Hada}, {Haggard}, {Haworth}, {Hecht}, {Hesper}, {Heumann}, {Ho}, {Ho}, {Honma}, {Huang}, {Huang}, {Hughes}, {Ikeda}, {Impellizzeri}, {Inoue}, {Issaoun}, {James}, {Jannuzi}, {Janssen}, {Jeter}, {Jiang}, {Jim{\'e}nez-Rosales}, {Johnson}, {Jorstad}, {Joshi}, {Jung}, {Karami}, {Karuppusamy}, {Kawashima}, {Keating}, {Kettenis}, {Kim}, {Kim}, {Kim}, {Kim}, {Kino}, {Koay}, {Kocherlakota}, {Kofuji}, {Koch}, {Koyama}, {Kramer}, {Kramer}, {Kramer}, {Krichbaum}, {Kuo}, {La Bella}, {Lauer}, {Lee}, {Lee}, {Leung}, {Levis}, {Li}, {Lico}, {Lindahl}, {Lindqvist}, {Lisakov}, {Liu}, {Liu}, {Liuzzo}, {Lo}, {Lobanov}, {Loinard}, {Lonsdale}, {Lowitz}, {Lu}, {MacDonald}, {Mao}, {Marchili}, {Markoff}, {Marrone}, {Marscher}, {Mart{\'\i}-Vidal}, {Matsushita}, {Matthews}, {Medeiros}, {Menten}, {Michalik}, {Mizuno}, {Mizuno}, {Moran}, {Moriyama}, {Moscibrodzka}, {Mulaudzi}, {M{\"u}ller}, {M{\"u}ller}, {Mus}, {Musoke}, {Myserlis}, {Nadolski}, {Nagai}, {Nagar}, {Nakamura}, {Narayanan}, {Natarajan},
      {Nathanail}, {Fuentes}, {Neilsen}, {Neri}, {Ni}, {Noutsos}, {Nowak}, {Oh}, {Okino}, {Olivares}, {Ortiz-Le{\'o}n}, {Oyama}, {{\"O}zel}, {Palumbo}, {Paraschos}, {Park}, {Parsons}, {Patel}, \& {Pen}}]{EHT_SgrA_P8}
    {Event Horizon Telescope Collaboration}, {Akiyama}, K., {Alberdi}, A., {et~al.} 2024{\natexlab{b}}, \apjl, 964, L26
    
    \bibitem[{{Event Horizon Telescope Collaboration} {et~al.}(2022{\natexlab{a}}){Event Horizon Telescope Collaboration}, {Akiyama}, {Alberdi}, {Alef}, {Algaba}, {Anantua}, {Asada}, {Azulay}, {Bach}, {Baczko}, {Ball}, {Balokovi{\'c}}, {Barrett}, {Baub{\"o}ck}, {Benson}, {Bintley}, {Blackburn}, {Blundell}, {Bouman}, {Bower}, {Boyce}, {Bremer}, {Brinkerink}, {Brissenden}, {Britzen}, {Broderick}, {Broguiere}, {Bronzwaer}, {Bustamante}, {Byun}, {Carlstrom}, {Ceccobello}, {Chael}, {Chan}, {Chatterjee}, {Chatterjee}, {Chen}, {Chen}, {Cheng}, {Cho}, {Christian}, {Conroy}, {Conway}, {Cordes}, {Crawford}, {Crew}, {Cruz-Osorio}, {Cui}, {Davelaar}, {De Laurentis}, {Deane}, {Dempsey}, {Desvignes}, {Dexter}, {Dhruv}, {Doeleman}, {Dougal}, {Dzib}, {Eatough}, {Emami}, {Falcke}, {Farah}, {Fish}, {Fomalont}, {Ford}, {Fraga-Encinas}, {Freeman}, {Friberg}, {Fromm}, {Fuentes}, {Galison}, {Gammie}, {Garc{\'\i}a}, {Gentaz}, {Georgiev}, {Goddi}, {Gold}, {G{\'o}mez-Ruiz}, {G{\'o}mez}, {Gu}, {Gurwell}, {Hada}, {Haggard}, {Haworth},
      {Hecht}, {Hesper}, {Heumann}, {Ho}, {Ho}, {Honma}, {Huang}, {Huang}, {Hughes}, {Ikeda}, {Impellizzeri}, {Inoue}, {Issaoun}, {James}, {Jannuzi}, {Janssen}, {Jeter}, {Jiang}, {Jim{\'e}nez-Rosales}, {Johnson}, {Jorstad}, {Joshi}, {Jung}, {Karami}, {Karuppusamy}, {Kawashima}, {Keating}, {Kettenis}, {Kim}, {Kim}, {Kim}, {Kim}, {Kino}, {Koay}, {Kocherlakota}, {Kofuji}, {Koch}, {Koyama}, {Kramer}, {Kramer}, {Krichbaum}, {Kuo}, {La Bella}, {Lauer}, {Lee}, {Lee}, {Leung}, {Levis}, {Li}, {Lico}, {Lindahl}, {Lindqvist}, {Lisakov}, {Liu}, {Liu}, {Liuzzo}, {Lo}, {Lobanov}, {Loinard}, {Lonsdale}, {Lu}, {Mao}, {Marchili}, {Markoff}, {Marrone}, {Marscher}, {Mart{\'\i}-Vidal}, {Matsushita}, {Matthews}, {Medeiros}, {Menten}, {Michalik}, {Mizuno}, {Mizuno}, {Moran}, {Moriyama}, {Moscibrodzka}, {M{\"u}ller}, {Mus}, {Musoke}, {Myserlis}, {Nadolski}, {Nagai}, {Nagar}, {Nakamura}, {Narayan}, {Narayanan}, {Natarajan}, {Nathanail}, {Fuentes}, {Neilsen}, {Neri}, {Ni}, {Noutsos}, {Nowak}, {Oh}, {Okino}, {Olivares}, {Ortiz-Le{\'o}n},
      {Oyama}, {{\"O}zel}, {Palumbo}, {Paraschos}, {Park}, {Parsons}, {Patel}, {Pen}, {Pesce}, {Pi{\'e}tu}, {Plambeck}, {PopStefanija}, {Porth}, {P{\"o}tzl}, {Prather}, {Preciado-L{\'o}pez}, \& {Psaltis}}]{2022ApJ...930L..13E}
    {Event Horizon Telescope Collaboration}, {Akiyama}, K., {Alberdi}, A., {et~al.} 2022{\natexlab{a}}, \apjl, 930, L13
    
    \bibitem[{{Event Horizon Telescope Collaboration} {et~al.}(2022{\natexlab{b}}){Event Horizon Telescope Collaboration}, {Akiyama}, {Alberdi}, {Alef}, {Algaba}, {Anantua}, {Asada}, {Azulay}, {Bach}, {Baczko}, {Ball}, {Balokovi{\'c}}, {Barrett}, {Baub{\"o}ck}, {Benson}, {Bintley}, {Blackburn}, {Blundell}, {Bouman}, {Bower}, {Boyce}, {Bremer}, {Brinkerink}, {Brissenden}, {Britzen}, {Broderick}, {Broguiere}, {Bronzwaer}, {Bustamante}, {Byun}, {Carlstrom}, {Ceccobello}, {Chael}, {Chan}, {Chatterjee}, {Chatterjee}, {Chen}, {Chen}, {Cheng}, {Cho}, {Christian}, {Conroy}, {Conway}, {Cordes}, {Crawford}, {Crew}, {Cruz-Osorio}, {Cui}, {Davelaar}, {De Laurentis}, {Deane}, {Dempsey}, {Desvignes}, {Dexter}, {Dhruv}, {Doeleman}, {Dougal}, {Dzib}, {Eatough}, {Emami}, {Falcke}, {Farah}, {Fish}, {Fomalont}, {Ford}, {Fraga-Encinas}, {Freeman}, {Friberg}, {Fromm}, {Fuentes}, {Galison}, {Gammie}, {Garc{\'\i}a}, {Gentaz}, {Georgiev}, {Goddi}, {Gold}, {G{\'o}mez-Ruiz}, {G{\'o}mez}, {Gu}, {Gurwell}, {Hada}, {Haggard}, {Haworth},
      {Hecht}, {Hesper}, {Heumann}, {Ho}, {Ho}, {Honma}, {Huang}, {Huang}, {Hughes}, {Ikeda}, {Violette Impellizzeri}, {Inoue}, {Issaoun}, {James}, {Jannuzi}, {Janssen}, {Jeter}, {Jiang}, {Jim{\'e}nez-Rosales}, {Johnson}, {Jorstad}, {Joshi}, {Jung}, {Karami}, {Karuppusamy}, {Kawashima}, {Keating}, {Kettenis}, {Kim}, {Kim}, {Kim}, {Kim}, {Kino}, {Koay}, {Kocherlakota}, {Kofuji}, {Koch}, {Koyama}, {Kramer}, {Kramer}, {Krichbaum}, {Kuo}, {La Bella}, {Lauer}, {Lee}, {Lee}, {Leung}, {Levis}, {Li}, {Lico}, {Lindahl}, {Lindqvist}, {Lisakov}, {Liu}, {Liu}, {Liuzzo}, {Lo}, {Lobanov}, {Loinard}, {Lonsdale}, {Lu}, {Mao}, {Marchili}, {Markoff}, {Marrone}, {Marscher}, {Mart{\'\i}-Vidal}, {Matsushita}, {Matthews}, {Medeiros}, {Menten}, {Michalik}, {Mizuno}, {Mizuno}, {Moran}, {Moriyama}, {Moscibrodzka}, {M{\"u}ller}, {Mus}, {Musoke}, {Myserlis}, {Nadolski}, {Nagai}, {Nagar}, {Nakamura}, {Narayan}, {Narayanan}, {Natarajan}, {Nathanail}, {Navarro Fuentes}, {Neilsen}, {Neri}, {Ni}, {Noutsos}, {Nowak}, {Oh}, {Okino}, {Olivares},
      {Ortiz-Le{\'o}n}, {Oyama}, {{\"O}zel}, {Palumbo}, {Filippos Paraschos}, {Park}, {Parsons}, {Patel}, {Pen}, {Pesce}, {Pi{\'e}tu}, {Plambeck}, {PopStefanija}, {Porth}, {P{\"o}tzl}, {Prather}, {Preciado-L{\'o}pez}, \& {Psaltis}}]{EHTC_SgrA_p5}
    {Event Horizon Telescope Collaboration}, {Akiyama}, K., {Alberdi}, A., {et~al.} 2022{\natexlab{b}}, \apjl, 930, L16
    
    \bibitem[{{Event Horizon Telescope Collaboration} {et~al.}(2021){Event Horizon Telescope Collaboration}, {Akiyama}, {Algaba}, {Alberdi}, {Alef}, {Anantua}, {Asada}, {Azulay}, {Baczko}, {Ball}, {Balokovi{\'c}}, {Barrett}, {Benson}, {Bintley}, {Blackburn}, {Blundell}, {Boland}, {Bouman}, {Bower}, {Boyce}, {Bremer}, {Brinkerink}, {Brissenden}, {Britzen}, {Broderick}, {Broguiere}, {Bronzwaer}, {Byun}, {Carlstrom}, {Chael}, {Chan}, {Chatterjee}, {Chatterjee}, {Chen}, {Chen}, {Chesler}, {Cho}, {Christian}, {Conway}, {Cordes}, {Crawford}, {Crew}, {Cruz-Osorio}, {Cui}, {Davelaar}, {De Laurentis}, {Deane}, {Dempsey}, {Desvignes}, {Dexter}, {Doeleman}, {Eatough}, {Falcke}, {Farah}, {Fish}, {Fomalont}, {Ford}, {Fraga-Encinas}, {Freeman}, {Friberg}, {Fromm}, {Fuentes}, {Galison}, {Gammie}, {Garc{\'\i}a}, {Gentaz}, {Georgiev}, {Goddi}, {Gold}, {G{\'o}mez}, {G{\'o}mez-Ruiz}, {Gu}, {Gurwell}, {Hada}, {Haggard}, {Hecht}, {Hesper}, {Ho}, {Ho}, {Honma}, {Huang}, {Huang}, {Hughes}, {Ikeda}, {Inoue}, {Issaoun}, {James},
      {Jannuzi}, {Janssen}, {Jeter}, {Jiang}, {Jimenez-Rosales}, {Johnson}, {Jorstad}, {Jung}, {Karami}, {Karuppusamy}, {Kawashima}, {Keating}, {Kettenis}, {Kim}, {Kim}, {Kim}, {Kim}, {Kino}, {Koay}, {Kofuji}, {Koch}, {Koyama}, {Kramer}, {Kramer}, {Krichbaum}, {Kuo}, {Lauer}, {Lee}, {Levis}, {Li}, {Li}, {Lindqvist}, {Lico}, {Lindahl}, {Liu}, {Liu}, {Liuzzo}, {Lo}, {Lobanov}, {Loinard}, {Lonsdale}, {Lu}, {MacDonald}, {Mao}, {Marchili}, {Markoff}, {Marrone}, {Marscher}, {Mart{\'\i}-Vidal}, {Matsushita}, {Matthews}, {Medeiros}, {Menten}, {Mizuno}, {Mizuno}, {Moran}, {Moriyama}, {Moscibrodzka}, {M{\"u}ller}, {Musoke}, {Mej{\'\i}as}, {Michalik}, {Nadolski}, {Nagai}, {Nagar}, {Nakamura}, {Narayan}, {Narayanan}, {Natarajan}, {Nathanail}, {Neilsen}, {Neri}, {Ni}, {Noutsos}, {Nowak}, {Okino}, {Olivares}, {Ortiz-Le{\'o}n}, {Oyama}, {{\"O}zel}, {Palumbo}, {Park}, {Patel}, {Pen}, {Pesce}, {Pi{\'e}tu}, {Plambeck}, {PopStefanija}, {Porth}, {P{\"o}tzl}, {Prather}, {Preciado-L{\'o}pez}, {Psaltis}, {Pu}, {Ramakrishnan}, {Rao},
      {Rawlings}, {Raymond}, {Rezzolla}, {Ricarte}, {Ripperda}, {Roelofs}, {Rogers}, {Ros}, {Rose}, {Roshanineshat}, {Rottmann}, {Roy}, {Ruszczyk}, {Rygl}, {S{\'a}nchez}, {S{\'a}nchez-Arguelles}, \& {Sasada}}]{2021ApJ...910L..12E}
    {Event Horizon Telescope Collaboration}, {Akiyama}, K., {Algaba}, J.~C., {et~al.} 2021, \apjl, 910, L12
    
    \bibitem[{{Frolov} \& {Shoom}(2010)}]{2010PhRvD..82h4034F}
    {Frolov}, V.~P. \& {Shoom}, A.~A. 2010, \prd, 82, 084034
    
    \bibitem[{{Gelles} {et~al.}(2021){Gelles}, {Himwich}, {Johnson}, \& {Palumbo}}]{2021PhRvD.104d4060G}
    {Gelles}, Z., {Himwich}, E., {Johnson}, M.~D., \& {Palumbo}, D. C.~M. 2021, \prd, 104, 044060
    
    \bibitem[{{Goddi} {et~al.}(2021){Goddi}, {Mart{\'\i}-Vidal}, {Messias}, {Bower}, {Broderick}, {Dexter}, {Marrone}, {Moscibrodzka}, {Nagai}, {Algaba}, {Asada}, {Crew}, {G{\'o}mez}, {Impellizzeri}, {Janssen}, {Kadler}, {Krichbaum}, {Lico}, {Matthews}, {Nathanail}, {Ricarte}, {Ros}, {Younsi}, {Akiyama}, {Alberdi}, {Alef}, {Anantua}, {Azulay}, {Baczko}, {Ball}, {Balokovi{\'c}}, {Barrett}, {Benson}, {Bintley}, {Blackburn}, {Blundell}, {Boland}, {Bouman}, {Boyce}, {Bremer}, {Brinkerink}, {Brissenden}, {Britzen}, {Broguiere}, {Bronzwaer}, {Byun}, {Carlstrom}, {Chael}, {Chan}, {Chatterjee}, {Chatterjee}, {Chen}, {Chen}, {Chesler}, {Cho}, {Christian}, {Conway}, {Cordes}, {Crawford}, {Cruz-Osorio}, {Cui}, {Davelaar}, {De Laurentis}, {Deane}, {Dempsey}, {Desvignes}, {Doeleman}, {Eatough}, {Falcke}, {Farah}, {Fish}, {Fomalont}, {Ford}, {Fraga-Encinas}, {Freeman}, {Friberg}, {Fromm}, {Fuentes}, {Galison}, {Gammie}, {Garc{\'\i}a}, {Gentaz}, {Georgiev}, {Gold}, {G{\'o}mez-Ruiz}, {Gu}, {Gurwell}, {Hada}, {Haggard}, {Hecht},
      {Hesper}, {Ho}, {Ho}, {Honma}, {Huang}, {Huang}, {Hughes}, {Inoue}, {Issaoun}, {James}, {Jannuzi}, {Jeter}, {Jiang}, {Jimenez-Rosales}, {Johnson}, {Jorstad}, {Jung}, {Karami}, {Karuppusamy}, {Kawashima}, {Keating}, {Kettenis}, {Kim}, {Kim}, {Kim}, {Kim}, {Kino}, {Koay}, {Kofuji}, {Koch}, {Koyama}, {Kramer}, {Kramer}, {Kuo}, {Lauer}, {Lee}, {Levis}, {Li}, {Li}, {Lindqvist}, {Lindahl}, {Liu}, {Liu}, {Liuzzo}, {Lo}, {Lobanov}, {Loinard}, {Lonsdale}, {Lu}, {MacDonald}, {Mao}, {Marchili}, {Markoff}, {Marscher}, {Matsushita}, {Medeiros}, {Menten}, {Mizuno}, {Mizuno}, {Moran}, {Moriyama}, {M{\"u}ller}, {Musoke}, {Mej{\'\i}as}, {Nagar}, {Nakamura}, {Narayan}, {Narayanan}, {Natarajan}, {Neilsen}, {Neri}, {Ni}, {Noutsos}, {Nowak}, {Okino}, {Olivares}, {Ortiz-Le{\'o}n}, {Oyama}, {{\"O}zel}, {Palumbo}, {Park}, {Patel}, {Pen}, {Pesce}, {Pi{\'e}tu}, {Plambeck}, {PopStefanija}, {Porth}, {P{\"o}tzl}, {Prather}, {Preciado-L{\'o}pez}, {Psaltis}, {Pu}, {Ramakrishnan}, {Rao}, {Rawlings}, {Raymond}, {Rezzolla}, {Ripperda},
      {Roelofs}, {Rogers}, {Rose}, {Roshanineshat}, {Rottmann}, {Roy}, {Ruszczyk}, {Rygl}, {S{\'a}nchez}, {S{\'a}nchez-Arguelles}, \& {Sasada}}]{2021ApJ...910L..14G}
    {Goddi}, C., {Mart{\'\i}-Vidal}, I., {Messias}, H., {et~al.} 2021, \apjl, 910, L14
    
    \bibitem[{{Goldreich} \& {Julian}(1969)}]{1969ApJ...157..869G}
    {Goldreich}, P. \& {Julian}, W.~H. 1969, \apj, 157, 869
    
    \bibitem[{{Gravity Collaboration} {et~al.}(2023){Gravity Collaboration}, {Abuter}, {Aimar}, {Amaro Seoane}, {Amorim}, {Baub{\"o}ck}, {Berger}, {Bonnet}, {Bourdarot}, {Brandner}, {Cardoso}, {Cl{\'e}net}, {Davies}, {de Zeeuw}, {Dexter}, {Drescher}, {Eckart}, {Eisenhauer}, {Feuchtgruber}, {Finger}, {F{\"o}rster Schreiber}, {Foschi}, {Garcia}, {Gao}, {Gelles}, {Gendron}, {Genzel}, {Gillessen}, {Hartl}, {Haubois}, {Haussmann}, {Hei{\ss}el}, {Henning}, {Hippler}, {Horrobin}, {Jochum}, {Jocou}, {Kaufer}, {Kervella}, {Lacour}, {Lapeyr{\`e}re}, {Le Bouquin}, {L{\'e}na}, {Lutz}, {Mang}, {More}, {Ott}, {Paumard}, {Perraut}, {Perrin}, {Pfuhl}, {Rabien}, {Ribeiro}, {Sadun Bordoni}, {Scheithauer}, {Shangguan}, {Shimizu}, {Stadler}, {Straub}, {Straubmeier}, {Sturm}, {Tacconi}, {Vincent}, {von Fellenberg}, {Widmann}, {Wielgus}, {Wieprecht}, {Wiezorrek}, \& {Woillez}}]{2023A&A...677L..10G}
    {Gravity Collaboration}, {Abuter}, R., {Aimar}, N., {et~al.} 2023, \aap, 677, L10
    
    \bibitem[{{Gravity Collaboration} {et~al.}(2018){Gravity Collaboration}, {Abuter}, {Amorim}, {Baub{\"o}ck}, {Berger}, {Bonnet}, {Brandner}, {Cl{\'e}net}, {Coud{\'e} Du Foresto}, {de Zeeuw}, {Deen}, {Dexter}, {Duvert}, {Eckart}, {Eisenhauer}, {F{\"o}rster Schreiber}, {Garcia}, {Gao}, {Gendron}, {Genzel}, {Gillessen}, {Guajardo}, {Habibi}, {Haubois}, {Henning}, {Hippler}, {Horrobin}, {Huber}, {Jim{\'e}nez-Rosales}, {Jocou}, {Kervella}, {Lacour}, {Lapeyr{\`e}re}, {Lazareff}, {Le Bouquin}, {L{\'e}na}, {Lippa}, {Ott}, {Panduro}, {Paumard}, {Perraut}, {Perrin}, {Pfuhl}, {Plewa}, {Rabien}, {Rodr{\'{\i}}guez-Coira}, {Rousset}, {Sternberg}, {Straub}, {Straubmeier}, {Sturm}, {Tacconi}, {Vincent}, {von Fellenberg}, {Waisberg}, {Widmann}, {Wieprecht}, {Wiezorrek}, {Woillez}, \& {Yazici}}]{2018A&A...618L..10G}
    {Gravity Collaboration}, {Abuter}, R., {Amorim}, A., {et~al.} 2018, \aap, 618, L10
    
    \bibitem[{{GRAVITY Collaboration} {et~al.}(2020{\natexlab{a}}){GRAVITY Collaboration}, {Baub{\"o}ck}, {Dexter}, {Abuter}, {Amorim}, {Berger}, {Bonnet}, {Brandner}, {Cl{\'e}net}, {Coud{\'e} Du Foresto}, {de Zeeuw}, {Duvert}, {Eckart}, {Eisenhauer}, {F{\"o}rster Schreiber}, {Gao}, {Garcia}, {Gendron}, {Genzel}, {Gerhard}, {Gillessen}, {Habibi}, {Haubois}, {Henning}, {Hippler}, {Horrobin}, {Jim{\'e}nez-Rosales}, {Jocou}, {Kervella}, {Lacour}, {Lapeyr{\`e}re}, {Le Bouquin}, {L{\'e}na}, {Ott}, {Paumard}, {Perraut}, {Perrin}, {Pfuhl}, {Rabien}, {Rodriguez Coira}, {Rousset}, {Scheithauer}, {Stadler}, {Sternberg}, {Straub}, {Straubmeier}, {Sturm}, {Tacconi}, {Vincent}, {von Fellenberg}, {Waisberg}, {Widmann}, {Wieprecht}, {Wiezorrek}, {Woillez}, \& {Yazici}}]{Michi2020}
    {GRAVITY Collaboration}, {Baub{\"o}ck}, M., {Dexter}, J., {et~al.} 2020{\natexlab{a}}, \aap, 635, A143
    
    \bibitem[{{GRAVITY Collaboration} {et~al.}(2020{\natexlab{b}}){GRAVITY Collaboration}, {Jim{\'e}nez-Rosales}, {Dexter}, {Widmann}, {Baub{\"o}ck}, {Abuter}, {Amorim}, {Berger}, {Bonnet}, {Brandner}, {Cl{\'e}net}, {de Zeeuw}, {Eckart}, {Eisenhauer}, {F{\"o}rster Schreiber}, {Garcia}, {Gao}, {Gendron}, {Genzel}, {Gillessen}, {Habibi}, {Haubois}, {Hei{\ss}el}, {Henning}, {Hippler}, {Horrobin}, {Jochum}, {Jocou}, {Kaufer}, {Kervella}, {Lacour}, {Lapeyr{\`e}re}, {Le Bouquin}, {L{\'e}na}, {Nowak}, {Ott}, {Paumard}, {Perraut}, {Perrin}, {Pfuhl}, {Rodr{\'\i}guez-Coira}, {Shangguan}, {Scheithauer}, {Stadler}, {Straub}, {Straubmeier}, {Sturm}, {Tacconi}, {Vincent}, {von Fellenberg}, {Waisberg}, {Wieprecht}, {Wiezorrek}, {Woillez}, {Yazici}, \& {Zins}}]{Alejandra2020}
    {GRAVITY Collaboration}, {Jim{\'e}nez-Rosales}, A., {Dexter}, J., {et~al.} 2020{\natexlab{b}}, \aap, 643, A56
    
    \bibitem[{Hazeltine \& Waelbroeck(2004)}]{HazeltineWaelbroeck2004}
    Hazeltine, R.~D. \& Waelbroeck, F.~L. 2004, The Framework of Plasma Physics, 1st edn. (Boca Raton: CRC Press), 344
    
    \bibitem[{{Jacquemin-Ide} {et~al.}(2024){Jacquemin-Ide}, {Rincon}, {Tchekhovskoy}, \& {Liska}}]{2024MNRAS.532.1522J}
    {Jacquemin-Ide}, J., {Rincon}, F., {Tchekhovskoy}, A., \& {Liska}, M. 2024, \mnras, 532, 1522
    
    \bibitem[{{Johnson} {et~al.}(2015){Johnson}, {Fish}, {Doeleman}, {Marrone}, {Plambeck}, {Wardle}, {Akiyama}, {Asada}, {Beaudoin}, {Blackburn}, {Blundell}, {Bower}, {Brinkerink}, {Broderick}, {Cappallo}, {Chael}, {Crew}, {Dexter}, {Dexter}, {Freund}, {Friberg}, {Gold}, {Gurwell}, {Ho}, {Honma}, {Inoue}, {Kosowsky}, {Krichbaum}, {Lamb}, {Loeb}, {Lu}, {MacMahon}, {McKinney}, {Moran}, {Narayan}, {Primiani}, {Psaltis}, {Rogers}, {Rosenfeld}, {SooHoo}, {Tilanus}, {Titus}, {Vertatschitsch}, {Weintroub}, {Wright}, {Young}, {Zensus}, \& {Ziurys}}]{2015Sci...350.1242J}
    {Johnson}, M.~D., {Fish}, V.~L., {Doeleman}, S.~S., {et~al.} 2015, Science, 350, 1242
    
    \bibitem[{{Kolo{\v{s}}} {et~al.}(2015){Kolo{\v{s}}}, {Stuchl{\'\i}k}, \& {Tursunov}}]{2015CQGra..32p5009K}
    {Kolo{\v{s}}}, M., {Stuchl{\'\i}k}, Z., \& {Tursunov}, A. 2015, Classical and Quantum Gravity, 32, 165009
    
    \bibitem[{{Komissarov}(2022)}]{2022MNRAS.512.2798K}
    {Komissarov}, S.~S. 2022, \mnras, 512, 2798
    
    \bibitem[{{Levis} {et~al.}(2024){Levis}, {Chael}, {Bouman}, {Wielgus}, \& {Srinivasan}}]{Levis2024}
    {Levis}, A., {Chael}, A.~A., {Bouman}, K.~L., {Wielgus}, M., \& {Srinivasan}, P.~P. 2024, Nature Astronomy, 8, 765
    
    \bibitem[{{Muslimov} \& {Tsygan}(1992)}]{1992MNRAS.255...61M}
    {Muslimov}, A.~G. \& {Tsygan}, A.~I. 1992, \mnras, 255, 61
    
    \bibitem[{{Narayan} {et~al.}(2021){Narayan}, {Palumbo}, {Johnson}, {Gelles}, {Himwich}, {Chang}, {Ricarte}, {Dexter}, {Gammie}, {Chael}, {Event Horizon Telescope Collaboration}, {Akiyama}, {Alberdi}, {Alef}, {Algaba}, {Anantua}, {Asada}, {Azulay}, {Baczko}, {Ball}, {Balokovi{\'c}}, {Barrett}, {Benson}, {Bintley}, {Blackburn}, {Blundell}, {Boland}, {Bouman}, {Bower}, {Boyce}, {Bremer}, {Brinkerink}, {Brissenden}, {Britzen}, {Broderick}, {Broguiere}, {Bronzwaer}, {Byun}, {Carlstrom}, {Chan}, {Chatterjee}, {Chatterjee}, {Chen}, {Chen}, {Chesler}, {Cho}, {Christian}, {Conway}, {Cordes}, {Crawford}, {Crew}, {Cruz-Osorio}, {Cui}, {Davelaar}, {De Laurentis}, {Deane}, {Dempsey}, {Desvignes}, {Doeleman}, {Eatough}, {Falcke}, {Farah}, {Fish}, {Fomalont}, {Ford}, {Fraga-Encinas}, {Friberg}, {Fromm}, {Fuentes}, {Galison}, {Garc{\'\i}a}, {Gentaz}, {Georgiev}, {Goddi}, {Gold}, {G{\'o}mez}, {G{\'o}mez-Ruiz}, {Gu}, {Gurwell}, {Hada}, {Haggard}, {Hecht}, {Hesper}, {Ho}, {Ho}, {Honma}, {Huang}, {Huang}, {Hughes}, {Ikeda},
      {Inoue}, {Issaoun}, {James}, {Jannuzi}, {Janssen}, {Jeter}, {Jiang}, {Jimenez-Rosales}, {Jorstad}, {Jung}, {Karami}, {Karuppusamy}, {Kawashima}, {Keating}, {Kettenis}, {Kim}, {Kim}, {Kim}, {Kim}, {Kino}, {Koay}, {Kofuji}, {Koch}, {Koyama}, {Kramer}, {Kramer}, {Krichbaum}, {Kuo}, {Lauer}, {Lee}, {Levis}, {Li}, {Li}, {Lindqvist}, {Lico}, {Lindahl}, {Liu}, {Liu}, {Liuzzo}, {Lo}, {Lobanov}, {Loinard}, {Lonsdale}, {Lu}, {MacDonald}, {Mao}, {Marchili}, {Markoff}, {Marrone}, {Marscher}, {Mart{\'\i}-Vidal}, {Matsushita}, {Matthews}, {Medeiros}, {Menten}, {Mizuno}, {Mizuno}, {Moran}, {Moriyama}, {Moscibrodzka}, {M{\"u}ller}, {Musoke}, {Mej{\'\i}as}, {Nagai}, {Nagar}, {Nakamura}, {Narayanan}, {Natarajan}, {Nathanail}, {Neilsen}, {Neri}, {Ni}, {Noutsos}, {Nowak}, {Okino}, {Olivares}, {Ortiz-Le{\'o}n}, {Oyama}, {{\"O}zel}, {Park}, {Patel}, {Pen}, {Pesce}, {Pi{\'e}tu}, {Plambeck}, {PopStefanija}, {Porth}, {P{\"o}tzl}, {Prather}, {Preciado-L{\'o}pez}, {Psaltis}, {Pu}, {Ramakrishnan}, {Rao}, {Rawlings}, {Raymond},
      {Rezzolla}, {Ripperda}, {Roelofs}, {Rogers}, {Ros}, {Rose}, {Roshanineshat}, {Rottmann}, {Roy}, {Ruszczyk}, {Rygl}, {S{\'a}nchez}, {S{\'a}nchez-Arguelles}, \& {Sasada}}]{2021ApJ...912...35N}
    {Narayan}, R., {Palumbo}, D. C.~M., {Johnson}, M.~D., {et~al.} 2021, \apj, 912, 35
    
    \bibitem[{{Ruffini} \& {Wilson}(1975)}]{1975PhRvD..12.2959R}
    {Ruffini}, R. \& {Wilson}, J.~R. 1975, \prd, 12, 2959
    
    \bibitem[{{Thorne} \& {MacDonald}(1982)}]{1982MNRAS.198..339T}
    {Thorne}, K.~S. \& {MacDonald}, D. 1982, \mnras, 198, 339
    
    \bibitem[{{Tursunov} \& {Dadhich}(2019)}]{2019Univ....5..125T}
    {Tursunov}, A. \& {Dadhich}, N. 2019, Universe, 5, 125
    
    \bibitem[{{Tursunov} {et~al.}(2016){Tursunov}, {Stuchl{\'{\i}}k}, \& {Kolo{\v s}}}]{2016PhRvD..93h4012T}
    {Tursunov}, A., {Stuchl{\'{\i}}k}, Z., \& {Kolo{\v s}}, M. 2016, \prd, 93, 084012
    
    \bibitem[{{Tursunov} {et~al.}(2020){Tursunov}, {Zaja{\v{c}}ek}, {Eckart}, {Kolo{\v{s}}}, {Britzen}, {Stuchl{\'\i}k}, {Czerny}, \& {Karas}}]{2020ApJ...897...99T}
    {Tursunov}, A., {Zaja{\v{c}}ek}, M., {Eckart}, A., {et~al.} 2020, \apj, 897, 99
    
    \bibitem[{{Vincent} {et~al.}(2024){Vincent}, {Wielgus}, {Aimar}, {Paumard}, \& {Perrin}}]{Vincent2024}
    {Vincent}, F.~H., {Wielgus}, M., {Aimar}, N., {Paumard}, T., \& {Perrin}, G. 2024, \aap, 684, A194
    
    \bibitem[{{Vos} {et~al.}(2022){Vos}, {Mo{\'s}cibrodzka}, \& {Wielgus}}]{2022A&A...668A.185V}
    {Vos}, J., {Mo{\'s}cibrodzka}, M.~A., \& {Wielgus}, M. 2022, \aap, 668, A185
    
    \bibitem[{{Wald}(1974)}]{1974PhRvD..10.1680W}
    {Wald}, R.~M. 1974, \prd, 10, 1680
    
    \bibitem[{{Wald}(1984)}]{1984ucp..book.....W}
    {Wald}, R.~M. 1984, {General Relativity}
    
    \bibitem[{{Walker} \& {Penrose}(1970)}]{1970CMaPh..18..265W}
    {Walker}, M. \& {Penrose}, R. 1970, Communications in Mathematical Physics, 18, 265
    
    \bibitem[{{Wielgus} {et~al.}(2024){Wielgus}, {Issaoun}, {Mart{\'\i}-Vidal}, {Emami}, {Moscibrodzka}, {Brinkerink}, {Goddi}, \& {Fomalont}}]{2024A&A...682A..97W}
    {Wielgus}, M., {Issaoun}, S., {Mart{\'\i}-Vidal}, I., {et~al.} 2024, \aap, 682, A97
    
    \bibitem[{{Wielgus} {et~al.}(2022{\natexlab{a}}){Wielgus}, {Marchili}, {Mart{\'\i}-Vidal}, {Keating}, {Ramakrishnan}, {Tiede}, {Fomalont}, {Issaoun}, {Neilsen}, {Nowak}, {Blackburn}, {Gammie}, {Goddi}, {Haggard}, {Lee}, {Moscibrodzka}, {Tetarenko}, {Bower}, {Chan}, {Chatterjee}, {Chesler}, {Dexter}, {Doeleman}, {Georgiev}, {Gurwell}, {Johnson}, {Marrone}, {Mus}, {Psaltis}, {Ripperda}, {Witzel}, {Akiyama}, {Alberdi}, {Alef}, {Carlos Algaba}, {Anantua}, {Asada}, {Azulay}, {Bach}, {Baczko}, {Ball}, {Balokovi{\'c}}, {Barrett}, {Baub{\"o}ck}, {Benson}, {Bintley}, {Blundell}, {Boland}, {Bouman}, {Boyce}, {Bremer}, {Brinkerink}, {Brissenden}, {Britzen}, {Broderick}, {Broguiere}, {Bronzwaer}, {Bustamante}, {Byun}, {Carlstrom}, {Ceccobello}, {Chael}, {Chatterjee}, {Chen}, {Chen}, {Cho}, {Christian}, {Conroy}, {Conway}, {Cordes}, {Crawford}, {Crew}, {Cruz-Osorio}, {Cui}, {Davelaar}, {De Laurentis}, {Deane}, {Dempsey}, {Desvignes}, {Dhruv}, {Dzib}, {Eatough}, {Emami}, {Falcke}, {Farah}, {Fish}, {Alyson Ford},
      {Fraga-Encinas}, {Freeman}, {Friberg}, {Fromm}, {Fuentes}, {Galison}, {Garc{\'\i}a}, {Gentaz}, {Gold}, {G{\'o}mez-Ruiz}, {G{\'o}mez}, {Gu}, {Hada}, {Haworth}, {Hecht}, {Hesper}, {Ho}, {Ho}, {Honma}, {Huang}, {Huang}, {Hughes}, {Ikeda}, {Violette Impellizzeri}, {Inoue}, {James}, {Jannuzi}, {Janssen}, {Jeter}, {Jiang}, {Jim{\'e}nez-Rosales}, {Jorstad}, {Joshi}, {Jung}, {Karami}, {Karuppusamy}, {Kawashima}, {Kettenis}, {Kim}, {Kim}, {Kim}, {Kim}, {Kino}, {Yi Koay}, {Kocherlakota}, {Kofuji}, {Koch}, {Koyama}, {Kramer}, {Kramer}, {Krichbaum}, {Kuo}, {La Bella}, {Lauer}, {Lee}, {Kin Leung}, {Levis}, {Li}, {Lico}, {Lindahl}, {Lindqvist}, {Lisakov}, {Liu}, {Liu}, {Liuzzo}, {Lo}, {Lobanov}, {Loinard}, {Lonsdale}, {Lu}, {Mao}, {Markoff}, {Marscher}, {Matsushita}, {Matthews}, {Medeiros}, {Menten}, {Michalik}, {Mizuno}, {Mizuno}, {Moran}, {Moriyama}, {M{\"u}ller}, {Musoke}, {Myserlis}, {Nadolski}, {Nagai}, {Nagar}, {Nakamura}, {Narayan}, {Narayanan}, {Natarajan}, {Nathanail}, {Navarro Fuentes}, {Neri}, {Ni}, {Noutsos},
      {Oh}, {Okino}, {Olivares}, {Ortiz-Le{\'o}n}, {Oyama}, {{\"O}zel}, {Palumbo}, {Filippos Paraschos}, {Park}, {Parsons}, {Patel}, {Pen}, {Pesce}, {Pi{\'e}tu}, {Plambeck}, \& {PopStefanija}}]{Wielgus2022_lightcurves}
    {Wielgus}, M., {Marchili}, N., {Mart{\'\i}-Vidal}, I., {et~al.} 2022{\natexlab{a}}, \apjl, 930, L19
    
    \bibitem[{{Wielgus} {et~al.}(2022{\natexlab{b}}){Wielgus}, {Moscibrodzka}, {Vos}, {Gelles}, {Mart{\'\i}-Vidal}, {Farah}, {Marchili}, {Goddi}, \& {Messias}}]{2022A&A...665L...6W}
    {Wielgus}, M., {Moscibrodzka}, M., {Vos}, J., {et~al.} 2022{\natexlab{b}}, \aap, 665, L6
    
    \bibitem[{{Yfantis} {et~al.}(2024{\natexlab{a}}){Yfantis}, {Mo{\'s}cibrodzka}, {Wielgus}, {Vos}, \& {Jimenez-Rosales}}]{2024A&A...685A.142Y}
    {Yfantis}, A.~I., {Mo{\'s}cibrodzka}, M.~A., {Wielgus}, M., {Vos}, J.~T., \& {Jimenez-Rosales}, A. 2024{\natexlab{a}}, \aap, 685, A142
    
    \bibitem[{{Yfantis} {et~al.}(2024{\natexlab{b}}){Yfantis}, {Wielgus}, \& {Mo{\'s}cibrodzka}}]{2024A&A...691A.327Y}
    {Yfantis}, A.~I., {Wielgus}, M., \& {Mo{\'s}cibrodzka}, M. 2024{\natexlab{b}}, \aap, 691, A327
    
    \bibitem[{{Yuan} {et~al.}(2003){Yuan}, {Quataert}, \& {Narayan}}]{2003ApJ...598..301Y}
    {Yuan}, F., {Quataert}, E., \& {Narayan}, R. 2003, \apj, 598, 301
    
    \bibitem[{{Zaja{\v c}ek} {et~al.}(2018){Zaja{\v c}ek}, {Tursunov}, {Eckart}, \& {Britzen}}]{2018MNRAS.480.4408Z}
    {Zaja{\v c}ek}, M., {Tursunov}, A., {Eckart}, A., \& {Britzen}, S. 2018, \mnras, 480, 4408
    
    \end{thebibliography}

\end{document}